\documentclass[reprint,amsmath,amssymb,aps,superscriptaddress,twocolumn,prl]{revtex4-1}

\usepackage{graphicx}% Include figure files
\usepackage{bm}% bold math
\usepackage[normalem]{ulem}
\usepackage{braket}

\usepackage{mathrsfs}
\usepackage[colorlinks]{hyperref}
\usepackage{color}

\begin{document}

\title{Giant non-linear susceptibility of hydrogenic donors in silicon and germanium} 

\author{Nguyen H. Le}
\affiliation{ Advanced Technology Institute and SEPNet, University of Surrey, Guildford, GU2 7XH, United Kingdom}
\author{Grigory V. Lanskii}
\affiliation{Institute of Monitoring of Climatic and Ecological Systems SB RAS, 10/3 Academical Ave., Tomsk 634055, Russia}
\author{Gabriel Aeppli}
\affiliation{ Laboratory for Solid State Physics, ETH Zurich, Zurich, CH-8093,Switzerland}
\affiliation{  Institut de Physique, EPF Lausanne, Lausanne, CH-1015, Switzerland} 
\affiliation{ Paul Scherrer Institut, Villigen PSI, CH-5232, Switzerland}
\author{Benedict N. Murdin}
\affiliation{ Advanced Technology Institute and SEPNet, University of Surrey, Guildford, GU2 7XH, United Kingdom}

\date{\today}

\begin{abstract}
Implicit summation is a technique for the conversion of sums over intermediate states in multiphoton absorption and the high-order susceptibility in hydrogen into simple integrals. Here, we derive the equivalent technique for hydrogenic impurities in multi-valley semiconductors. While the absorption has useful applications, it is primarily a loss process; conversely, the non-linear susceptibility is a crucial parameter for active photonic devices. For Si:P, we predict the hyperpolarizability ranges from $\chi^{(3)}/n_{\text{3D}}=2.9 $ to $580 \times 10^{-38}$ $\text{m}^5/\text{V}^2$ depending on the frequency, even while avoiding resonance. Using samples of a reasonable density, $n_{\text{3D}}$, and thickness, $L$, to produce third-harmonic generation at 9 THz, a frequency that is difficult to produce with existing solid-state sources, we predict that $\chi^{(3)}$ should exceed that of bulk InSb and $\chi^{(3)}L$ should exceed that of graphene and resonantly enhanced quantum wells.
\end{abstract}

\maketitle

\section{Introduction}
Multiphoton absorption requires a high intensity, and was first observed shortly after the invention of the laser using impurities in solids \cite{Kaiser1961} and alkali vapour \cite{Abella1962}. Although multiphoton absorption is useful for metrology and modulators, and can be enhanced where there is near-resonance of an intermediate state as in the case of Rb \cite{Saha}, it is essentially a loss process contributing an imaginary part to the non-linear susceptibility. The corresponding real part is responsible for a great variety of wavelength conversion processes such as harmonic generation, first observed in quartz \cite{Franken1961} and later in atomic vapours \cite{Ward1969} including alkalis \cite{MH}. 
THz multiphoton absorption has been shown to be very large in hydrogenic shallow impurities in semiconductors, even without intermediate state resonances \cite{vanLoon2018}, due to the large dielectric screening and low effective mass. Here, we predict giant values for the real part of the THz non-linear susceptibility for doped silicon and germanium. This finding opens access to novel applications for these materials in THz photonics. For example, tripling the output of a 2-4 THz quantum cascade laser through third-harmonic generation would fill the frequency gap currently only filled by larger, more expensive systems. We show that a good efficiency can be obtained for third-harmonic generation with doped silicon and germanium. Our theory can be readily applied to any donor in any semiconductor host where the effective mass approximation is valid, and our discussion makes it clear that a giant value of $\chi^{(3)}$ is expected for donors with a small binding energy in a host with a large dielectric constant and small effective mass.

The theory developed in this paper is appropriate for frequencies both near to and far from loss-inducing resonances, including the effects of effective mass anisotropy, multi-valley interactions and the central cell correction. The method could easily be applied to other systems with complicated potentials, such as multi-quantum wells.  Although this work focuses on perturbative harmonic generation, we anticipate that shallow impurities may also be useful for non-perturbative high-harmonic generation (HHG) \cite{Vampa2014,Beaulieu2016} taking advantage of the excellent control over the carrier-envelope phase of few-cycle pulses in this THz regime, which can be used to enhance HHG \cite{Haworth2007}.

\section{Results}
\subsection{The implicit summation technique}
From \textit{N}th-order perturbation theory  \cite{vanLoon2018, citeGT} the N-photon absorption (NPA) transition rate may be written as 
\begin{equation}\label{rate}
w^{(N)}=2 \pi \frac{ \left( 2 \pi  \alpha _{fs} \right) ^{N}}{N} \vert M^{(N)} \vert ^2  \left[ \frac{E_H^2}{ \varepsilon _r^{N/2}I_a^N} \right] \frac{I_m^N \Gamma^{(N)}}{\hbar^2},
\end{equation}
where $ I_a=E_H^2/\hbar a_B^2 $, $ a_B$ is the Bohr radius, $E_H$  the Hartree energy, and $  \alpha _{fs} $   the fine structure constant. $ M^{(N)} $ is a dimensionless transition matrix element,  and $I_m$  is the intensity of the light in the medium with relative dielectric permittivity  $  \varepsilon _r $. The lineshape function  $ \Gamma^{(N)} \left(  \omega  \right)  $  has unit area. For silicon and germanium donors, the factors inside the bracket are renormalized, and of particular importance here $I_a$ is ten orders of magnitude smaller for silicon than it is for hydrogen. This is apparent from the formulae of the Hartree energy and Bohr radius for donors in these materials: $E_H=m_t (e^2/4\pi \epsilon_0\epsilon\hbar)^2$, and $a_B=4\pi \epsilon_0\epsilon \hbar^2 /m_t e^2$, where $m_t$ is the transverse effective mass and $\epsilon$ the dielectric constant \cite{nicolePRB}. Both germanium and silicon have a small $m_t$ and large $\epsilon$, raising the Bohr radius and lowering the binding energy. The wavefunction is therefore significantly larger than that of alkali atoms, leading to an enhanced dipole matrix element and hence a substantially stronger interaction with light. 

The details of the spectrum given by Eqn(\ref{rate}) are controlled by  $ M^{(N)} $, which is influenced in silicon by the indirect valley structure, the anisotropic effective mass, and the donor central cell correction potential. Our main aim here is to calculate these effects. For single-photon absorption ($N=1$) between states  $ \ket{\psi_g} $  (the ground state) and  $ \ket{\psi_e} $  (the excited state),  
$
M^{(1)}= \bra{\psi_e}   \boldsymbol{\epsilon}.\bm{r}  \ket{\psi_g }  /a_B,
$
where $ \boldsymbol{\epsilon} $ is a unit vector in the polarization direction, and  Eqn (\ref{rate}) reduces to Fermi's golden rule. For two-photon absorption,
\begin{align}
M^{(2)}=\frac{E_H}{\hbar a_B^2} \sum_j\frac{\bra{\psi_e}\bm{\epsilon}.\bm{r} \ket{j}\bra{j} \bm{\epsilon}.\bm{r} \ket{\psi_g}}{\omega_{jg}-\omega_{eg}/2}, \nonumber
\end{align}
in the $\bm{E.r}$ gauge, which may be written as $M^{(2)}
= \bra{\psi_e} \zeta  G_1  \zeta \ket{\psi_g} $  
where $  \zeta = \boldsymbol{\epsilon}.\bm{r}/a_B $, 
\begin{equation}\label{multiphoton_operator}
G_n=\frac{E_H}{\hbar} \sum _j \frac{  \ket{ j }   \bra{ j }  }{ \left(\omega _{jg}- n \omega \right) },    
\end{equation}
and $\omega=\omega _{eg}/N$.  The states $   \ket{ j } $ are intermediate states, and along with $   \ket{\psi_e}  $ \& $  \ket{\psi_g}   $ they are eigenstates of $H  \ket{ j }  = \hbar \omega _j   \ket{ j }$,  
where $H$ is the Hamiltonian in the dark. For general multiphoton absorption, 
\begin{equation}\label{multiphoton_matrix}
M^{(N \geq 2)}
= \bra{\psi_e} \zeta G_{N-1}\zeta \ldots \zeta G_2 \zeta G_1  \zeta \ket{\psi_g}. 
\end{equation}

The summation in Eqn (\ref{multiphoton_operator})  can be avoided 
\cite{citeGT} by noticing that $\left(H- W_n \right)G_n= E_H$, where $W_n=\hbar \omega _g +n\hbar \omega$,
and $\omega=\omega_{eg}/N$ as already mentioned, and by using the completeness relation  $  \sum _j \ket{j} \bra{j} =1 $. In other words,
\begin{equation}\label{multiphoton_operator_implicit}
G_n=  E_H \left(H- W_n \right)^{-1}.
\end{equation}
Rewriting Eqn (\ref{multiphoton_matrix}), $M^{(N)}=\bra{\psi_e}  \zeta   \ket{\psi_{N-1}}  $ where $   \ket{\psi_0}  =  \ket{\psi_g}   $ and $ \ket{\psi_n}$ is the solution of the partial differential equation (PDE) $ G_n^{-1}\ket{\psi_n}  =  \zeta  \ket{ \psi_{n-1}}  $. 
Instead of finding $M^{(N)}$ by repeated application of Eqn (\ref{multiphoton_operator}), which requires infinite sums (that might be reduced down to a few terms if there are obvious resonances), we may now use Eqn (\ref{multiphoton_operator_implicit}) and the PDE at each stage, which  can be simpler.  

The Nth-order susceptibility far from any multiphoton resonances may also be calculated using the Nth-order perturbation theory  \cite{Boyd}. For example, the ``resonant" term in the third-order susceptibility, $\chi^{(3)}(3\omega)$, is 
\begin{align} 
 \frac{n_{\text{3D}}e^4}{\epsilon_0\hbar^3} & \sum_{l,k,j}  
  \frac{\bra{\psi_g}\bm{\epsilon}.\bm{r} \ket{l}\bra{l}\bm{\epsilon}.\bm{r} \ket{k}\bra{k}\bm{\epsilon}.\bm{r}  \ket{j}\bra{j}\bm{\epsilon}.\bm{r} \ket{\psi_g}}{(\omega_{lg}-3\omega)(\omega_{kg}-2\omega)(\omega_{jg}-\omega)},  \nonumber
 \end{align}
where $e$ is the electron charge, and $n_{\text{3D}}$ is the concentration. $\chi^{(3)}$ may be written in a similar form to Eqns (\ref{rate}) \& (\ref{multiphoton_matrix}), and for $N^{\text{th}}$ order,
\begin{equation}\label{chiN}
 \chi ^{(N)}=C^{(N)} \left[ \frac{a_B}{ I_a^{N/2}} \right] \frac{ n_{\text{3D}}e^{N+1}}{ \hbar^{N/2} \varepsilon _{0}},
\end{equation}
where $ C^{(N)}=\bra{\psi_g} \zeta G_{N} \ldots G_2 \zeta
G_1 \zeta\ket{\psi_g} $  is a dimensionless matrix element that may be found in a similar way to $M^{(N)}$, either by repeated application of Eqn (\ref{multiphoton_operator})--- as has been done previously for alkali metal vapours \cite {MH}--- or by using the implicit summation method of Eqn (\ref{multiphoton_operator_implicit}) with the only difference being $\omega \neq \omega_{eg}/N$. The antiresonant terms \cite{Boyd} and other non-linear processes, such as sum-frequency generation, can be calculated with simple modifications to $W_n$ at each step. 

\subsection{Multi-valley theory for donors in silicon and germanium}
In this section, we develop the multi-valley theory for the nonlinear optical processes of donors based on the effective mass approximation (EMA). For simplicity of presentation, we describe the derivation for silicon; the case of germanium is discussed in the Supplementary Materials. It will become apparent that our theory is readily applicable to any donor in any host as long as the EMA is reliable.

To apply the method to donors, we require $ \ket{\psi_g} , \omega_g,  \ket{\psi_e} , \omega_e $ and $H \ket{\psi_n} $. Silicon and germanium are indirect with equivalent conduction band minima (valleys) near the Brillouin zone edge; each minimum is characterized by a Fermi surface that is a prolate ellipsoid with transverse \& longitudinal effective masses,  $m_{t,l}$. According to the Kohn-Luttinger effective mass approximation \cite{KL1955}, the state $\ket{\psi_j}$ of a shallow donor can be decomposed into slowly varying hydrogenic envelope functions, one for each valley, modulated by plane-wave functions corresponding to the crystal momenta at the minima, $\bm{k}_{\mu} $ (and a lattice periodic function that is unimportant here).  We write 
$
 \psi_j(\bm{r})   = \sum _\mu   e^{  i\bm{k}_\mu .\bm{r} }  F_{j,\mu}(\bm{r}),
$
where $F_{j,\mu}(\bm{r})$ is the slowly varying envelope function. We have neglected the lattice periodic part, $u_{\mu}(\bm{r})$, of the Bloch functions for the simplicity of presentation. A rigorous derivation with $u_{\mu}(\bm{r})$ included is provided in the Supplementary Materials, but it does not lead to any change in the final equations for the envelope functions (Eqns~\eqref{eigenproblem} and \eqref{eq:coupledPDEs} below).

We separate the potential into the slowly varying Coulomb term of the donor $ V (\bm{r})$, and a rapidly varying term due to the quantum defect that is short range, $ U ( \bm{r} )  $, referred to as the central cell correction (CCC). Within the EMA, the kinetic energy term in the Hamiltonian operates only on the envelope function, and the EMA Schrodinger equation may be written as
\begin{equation}\label{EMA}
  \sum _\mu e^{  i\bm{k}_{ \mu }.\bm{r} }    \left[ H_0+U -\hbar\omega_j\right] F_{j,\mu}(\bm{r}) =0,
\end{equation}
where $H_0$ includes the Coulomb potential $V(\bm{r})$:
$ E_H^{-1}H_0=-\frac{1}{2}a_B^2\left[\partial_x^2+\partial_y^2+\gamma \partial_z^2\right]-a_B r^{-1} $
using a valley-specific coordinate system ($x,y,z$ where $z$ is the valley axis, i.e., the valley-frame is rotated relative to the lab-frame of $x_1,x_2,x_3$). The kinetic energy has cylindrical symmetry because $\gamma =m_t/m_l \neq 1$, and $V(\bm{r})$ and $U(\bm{r})$ are spherical and tetrahedral respectively. $H_0$ produces wavefunctions that are approximately hydrogen-like, and $U(\bm{r})$ mixes them to produce states that transform as the $\text{A}_1, \text{E}$ and $\text{T}_2$ components of the $\text{T}_d$ point group. 

We take  $ U ( \bm{r})  $ to be very short range, and we neglect the small change in the envelope functions over the short length scale  $2 \pi /\vert \bm{k}_{ \mu }\vert$.
Premultiplying Eqn (\ref{EMA}) by  $ e^{-i \bm{k}_{\mu'} . \bm{r} } $  and averaging over a volume  $ ( 2 \pi /\vert \bm{k}_{ \mu }\vert ) ^3 $ around $\bm{r}$, the Schrodinger eqn now reads 
$ 
\left[ H_0 - \hbar\omega_j  \right] F_{j,\mu}(\bm{r}) +\sum _{\mu'} U_{\mu\mu'}  \delta(\bm{r}) F_{j,\mu'}(\bm{r}) =0,
$
where  $\delta(\bm{r})$ is the Dirac delta function, and $U_{\mu \mu'}=\int d\bm{r}\, e^{i (\bm{k}_{\mu'}-\bm{k}_{\mu}).\bm{r}}U(\bm{r})$.  For an $\text{A}_{1}$ state, all the envelope functions have the same amplitude at $r=0$, hence,  $\sum _{\mu'} U_{\mu\mu'}  \delta(\bm{r}) F_{j,\mu'}(\bm{r}) =-U_{cc} \delta(\bm{r})  F_{j,\mu}(\bm{r}) $, where $U_{cc}=- \sum _{\mu'} U_{\mu\mu'}$. It is found experimentally that for $\text{E}$ and $\text{T}_2$ states, the CCC has a rather small effect, and so we neglect it. Since $H_0$ has cylindrical symmetry, the component of angular momentum about the valley axis is a conserved quantity, i.e., 
$
F_{j, \mu}(\bm{r})  = e^{im\phi} f_{j,m, \mu}(r,\theta),
$
where $m$ is a good quantum number, and now $f_{j,m,\mu}$ is a 2D function only. Substituting into the Schrodinger eqn, premultiplying by $e^{-im'\phi}$ and finally integrating over $\phi$, the eigenproblems are 
\begin{eqnarray}\label{eigenproblem}
 \left[H_0^{(m)}  - U_{cc} \delta (r)  -\hbar \omega _j \right] f_{j, m,\mu }^{(A_1)}(r,\theta)  =0, \nonumber \\
 \left[ H_0^{(m)} -\hbar \omega _{j} \right] f_{j,m, \mu}^{(E,T_2)}(r,\theta)  =0,
\end{eqnarray}
where 
$H_0^{(m)}=H_0+E_H a_B^2 m^2/2(r\sin\theta)^2$. We solve Eqns (\ref{eigenproblem}) using a 2D finite element method (FEM) (see Supplementary Materials). 
\begin{figure}[t]
\centering
\includegraphics[width=0.45\textwidth]{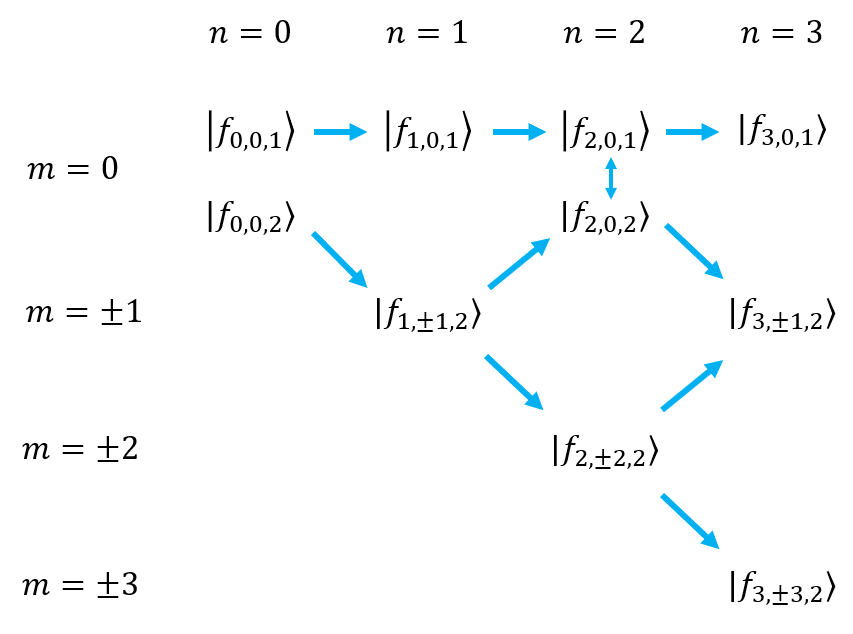}
\caption{\label{fig:interaction} Multiphoton intermediate states $\ket{f_{n,m,\mu}}$ and their interactions produced by dipole excitation polarized along $x_1$ (horizontal
arrows for the $\mu=1$ valley and diagonal arrows for the $\mu=2$ valley) and produced by $U_{cc}$ (vertical arrows).}
\end{figure}

We focus on silicon, in which case the valley index, $\mu$, runs over ($\pm 1,\pm 2, \pm 3$), where 1,2,3 are the three crystal axes, and we let the light be polarized along a crystal axis, $x_1 $, by way of illustration; the calculation for germanium and other polarization directions is described in the Supplementary Materials. For the $\mu=\pm 1,\pm 2,\pm 3$ valleys, $a_B \zeta_\mu=z,x,y=r\cos \theta,r\sin \theta \cos\phi, r\sin \theta \sin\phi$, respectively, because each  has its coordinate rotated so that $z$ is the valley axis. Following the expansion of $\psi_j$ in terms of the $f_{j,m, \mu}$, we write the intermediate state functions as $
 \psi_{n}(\bm{r}) = \sum_{m,\mu} e^{im\phi} e^{  i\bm{k}_\mu.\bm{r} }  f_{n,m,\mu }(r,\theta)
$, substitute them into $ G_n^{-1}\psi_n  =  \zeta   \psi_{n-1}  $, premultiply by  $ e^{ - i\bm{k}_{\mu'} . \bm{r} } $, average over a volume of $  ( 2 \pi /\vert \mathbf{k}_\mu  \vert)^3 $,  premultiply by $e^{-im'\phi}$, and finally, integrate over $\phi$. Since $f_{0,0,\mu } =f_{ g,0,\mu} $ for all $\mu$, we find that $f_{ n,m,3} = i^{-m}f_{ n,m,2 }$ and $f_{n,m,-\mu } =f_{n,m,\mu} $, and 

\begin{widetext}
\begin{eqnarray}\label{eq:coupledPDEs}
 \left[{H}_0^{(m)}-W_n- \mathcal{D}  \right] f_{n,m,1} -2  \mathcal{D} f_{n,m,2} &=&  (E_H/a_B) \,r\cos\theta f_{n-1,m,1}, \nonumber  \\ 
 \left[ {H}_0^{(m)}- W_n-2   \mathcal{D}  \right] f_{n,m,2}   -\mathcal{D}f_{n,m,1}   &=&  (E_H/a_B) \,r \sin\theta \left[  f_{n-1,m-1,2} + f_{n-1,m+1,2} \right]/2, \label{eqn:silicon_implicit_PDE}
\end{eqnarray}
\end{widetext}
where $ \mathcal{D} = U_{cc} \delta (\bm{r})\delta_{m,0} /3 $ and $\delta_{m,0}$ is the Kronecker delta. In the above equations we drop the valley-specific coordinates in $f_{n,m,\mu}$ for notational simplicity, and the coordinates in $H_0^{(m)}$ and the right hand side are understood to belong to the valley of the envelope function that they act on.

\begin{figure}[t]
\centering
\includegraphics[width=0.45\textwidth]{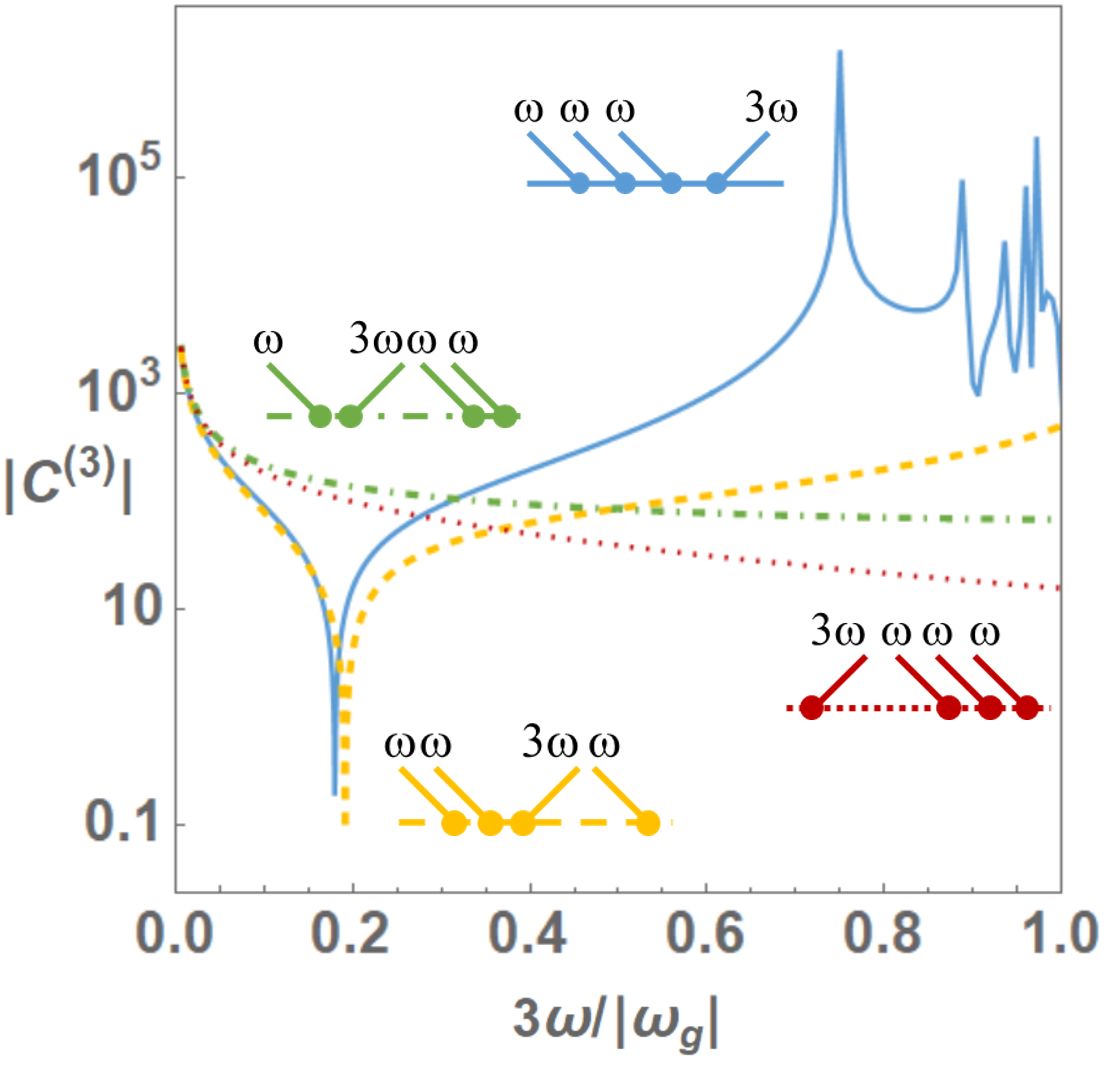}
\caption{\label{fig:antires} Contributions to $C^{(3)}$ from the resonant and anti-resonant terms for hydrogen: $\bra{\psi_0} \zeta G_3 \zeta G_2 \zeta G_1 \zeta\ket{\psi_0}$ (blue),
$\bra{\psi_0} \zeta G_{-1} \zeta G_2 \zeta G_1 \zeta\ket{\psi_0}$ (yellow),
$\bra{\psi_0} \zeta G_{-1} \zeta G_{-2} \zeta G_1 \zeta\ket{\psi_0}$ (green), and 
$\bra{\psi_0} \zeta G_{-1} \zeta G_{-2} \zeta G_{-3} \zeta\ket{\psi_0}$ (red). At $\omega=0$, the two terms containing $G_{-2}$ have opposite signs to the two terms with $G_2$, and the sum tends to 222.}
\end{figure}

It is evident that Eqns (\ref{eqn:silicon_implicit_PDE}) are not coupled by $ U_{cc} $  when the envelope function is zero at the origin. The ground state  $\ket{\psi_0}=\ket{\psi_g}$  has only  $ m=0 $ components, and it has even parity. Therefore,  $\ket{\psi_1}$  has odd parity according to Eqns (\ref{eqn:silicon_implicit_PDE}), so the  $ U_{cc} $  coupling term is suppressed.  By the same logic, the  $ U_{cc} $  coupling is only non-zero for even  $ n $ and $m=0$. In the case of  $  \ket{f_{n,m,1}} $, there is only dipole coupling to the functions with the same  $ m $, while for  $  \ket{f_{n,m,2}}$  the dipole coupling is to states with  $  \Delta m=~\pm~1 $. The latter couplings are identical, so $  f_{n,-m,\mu} =  f_{n,m,\mu} $. Figure \ref{fig:interaction} shows how the intermediate states are coupled by dipole excitation and the CCC. 

Eqns (\ref{eqn:silicon_implicit_PDE}) can be solved by sequential application of the 2D FEM \cite{Mathematica}. To test our numerical calculation we first compute $C^{(3)}$ for hydrogen, and each of the resonant and antiresonant terms is shown in Fig~\ref{fig:antires}. Their sum is shown in Fig~\ref{fig:chi3spectrum}, and we find excellent agreement within 0.2\% of the previous result obtained from a Sturmian coulomb Green function in Ref.~\cite{Mizuno1972}. 

\section{Discussion}
\subsection{Giant third-order nonlinear susceptibility}
\begin{figure}[t]
\centering
\includegraphics[width=0.45\textwidth]{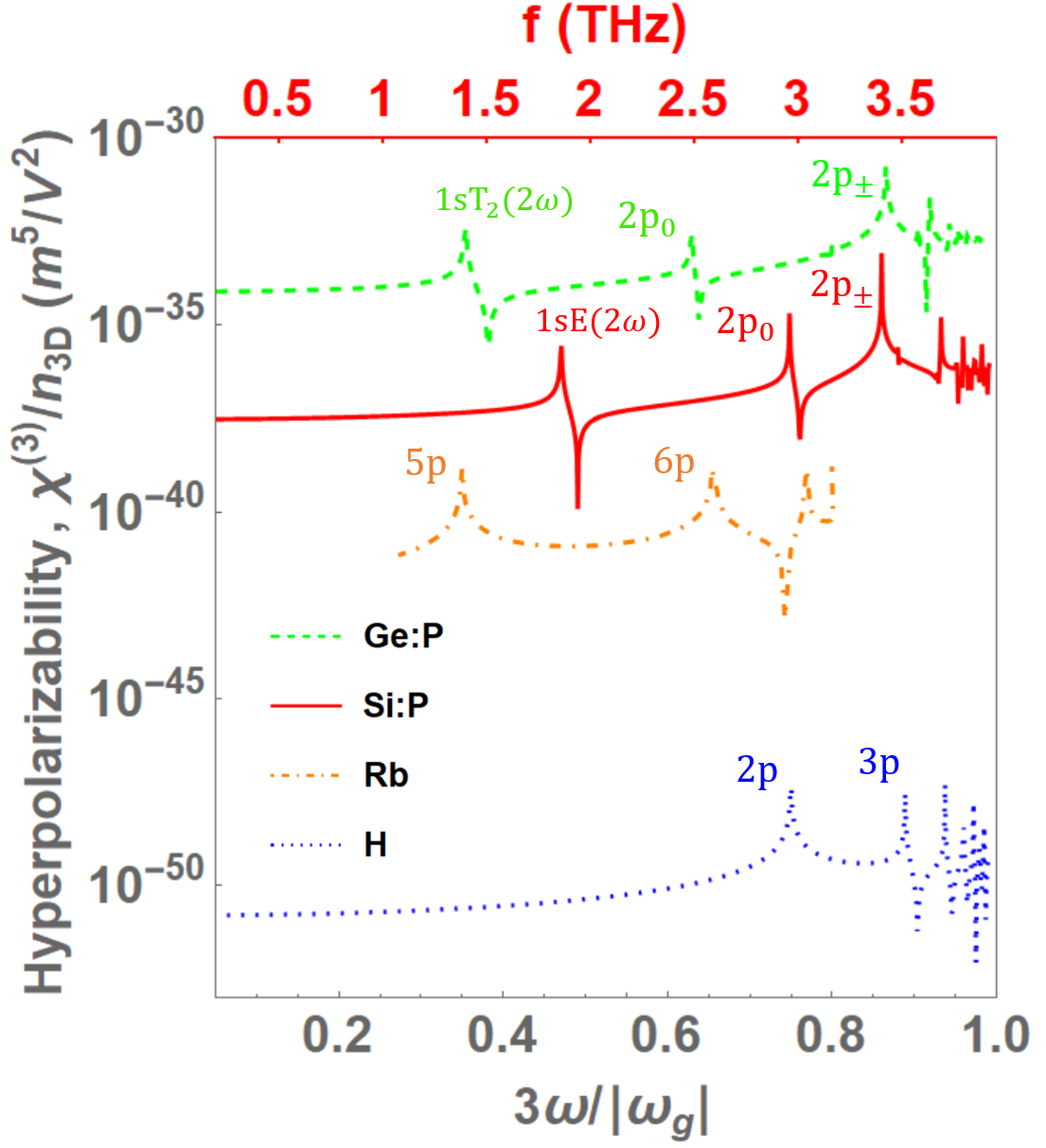}
\caption{\label{fig:chi3spectrum} The $\chi^{(3)}$ spectrum for hydrogenic donors Si:P and Ge:P, with light polarized along a valley axis in each case, and hydrogen (all calculations from this work). A hydrogenic atomic vapour (Rb) is shown for comparison (data from Ref.~\cite{MH}). Labels indicate the excited state for $3\omega=\omega_{eg}$ resonances and one $2\omega$ resonance. The top axis applies only to Si:P and indicates the frequency in THz.}
\end{figure}

\begin{figure}[b]
\centering
\includegraphics[width=0.45\textwidth]{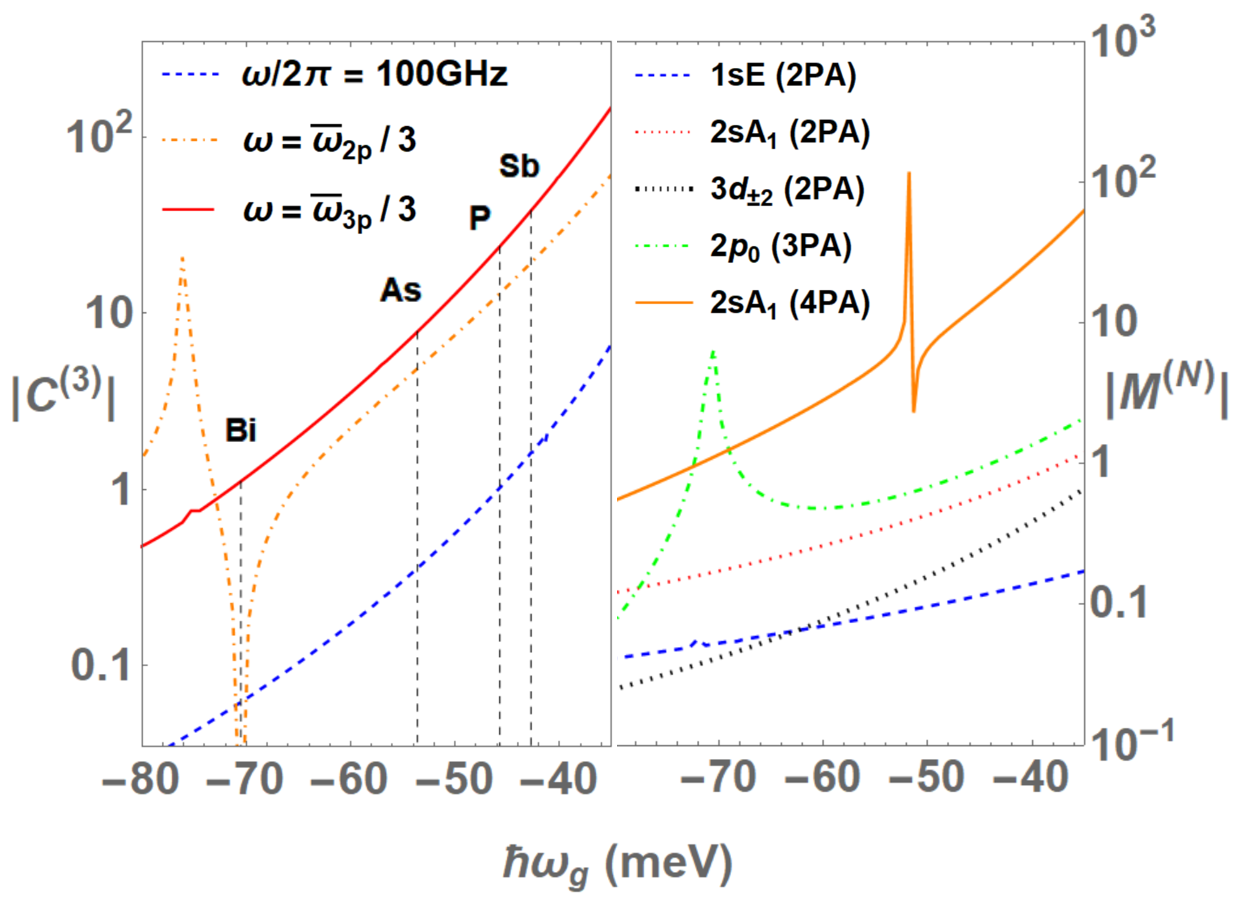}
\caption{\label{fig:Ucceffect} The effect of the CCC on $C^{(3)}$ (left panel) and the NPA absorption matrix element $M^{(N)}$ (right panel). The abscissa is the binding energy of the ground state (which is 31.5meV at $U_{cc}$=0), $\bar{\omega}_{2p}=(\omega_{2p_0}+\omega_{2p_{\pm}})/2-\omega_g$ is the average transition frequency to the $2p$ levels, and likewise for $\bar{\omega}_{3p}$. The binding energies of the Group V shallow donors are indicated in the left panel. The resonance and zero-crossing in the left panel, as well as the peaks in the $2p_0$ (3PA)  and $2s\text{A}_1$ (4PA) matrix element on the right are due to the  $(2\omega)$ resonance with the intermediate 1sE state.}
\end{figure}

\begin{figure}[t]
\centering
\includegraphics[width=0.45\textwidth]{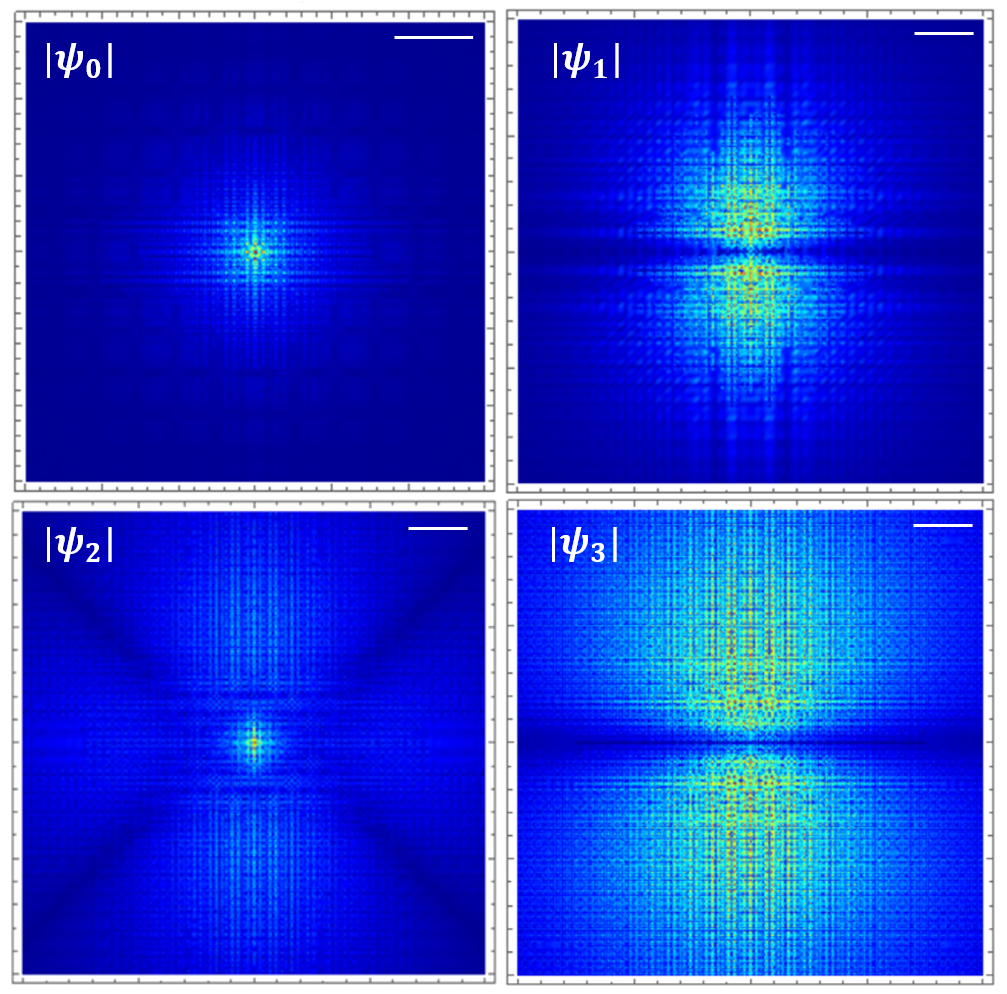}
\caption{\label{fig:wavefunctions} The wavefunctions $\ket{\psi_0}, \ket{\psi_1}, \ket{\psi_2}$ and $\ket{\psi_3}$ for Si:P (i.e., a binding energy of $\hbar \omega_g =-45.5$meV) in the $x_3=0$ plane. The frequency used for this calculation is the average of the $2p_0$ and $2p_\pm$ resonances, and the colour scale is normalized separately for each panel. The white bars on the top right indicate a length scale of 5nm.}
\end{figure}

Since silicon and germanium donors have an isotropic potential in an isotropic dielectric, the lowest-order nonlinear response is determined by $\chi^{(3)}$. The $\chi^{(3)}$  spectrum for each (including the antiresonant terms) is shown in Fig \ref{fig:chi3spectrum}. We took the parameters for silicon obtained from spectroscopic \cite{Ramdas1981} and magneto-optical measurements \cite{murdin2013,nicolePRB}, which are $\gamma \approx 0.208$, $a_B\approx  3.17$ nm and $E_H\approx 39.9$ meV. The parameters for germanium are $\gamma\approx 0.0513$, $a_B\approx 9.97$ nm and $E_H\approx 9.40$ meV \cite{Faulkner1969}. Resonances occur when $3\omega=\omega_{eg}$, labelled according to $\ket{\psi_e}$, and there are also sign-changes at which $\vert \chi^{(3)} \vert$ goes to zero. 
In the range of frequency shown, we also observe a two-photon resonance for  $1s \text{A}_1 \to 1s\text{E}$, which is an obvious illustration of the need for a multivalley theory. There is no $3\omega$ resonance with $1s \text{T}_2$ within the approximations made above in which   there is no intervalley dipole coupling. The effect of $U_{cc}$ on $\chi^{(3)}$ and the NPA matrix element is shown in Fig \ref{fig:Ucceffect}. The low-frequency response of $C^{(3)}$ is illustrated at 100 GHz. Two higher-frequency curves are included, with both far from $3\omega$ resonances, half way between the $2p_0$ and $2p_\pm$ resonances, and between the $3p_0$ and $3p_\pm$. We choose these average frequencies since $\chi^{(3)}$ for Si:P varies slowly around them (see Fig.~\ref{fig:chi3spectrum}) and hence would not be sensitive to small experimental variations in the light frequency. For the $2p$-average frequency, the $2\omega$ resonance with the $1s \text{E}$ produces a coincidental zero-crossing for Si:Bi. Example results for the intermediate state wavefunctions produced in the calculation are shown in Fig \ref{fig:wavefunctions}. The state $\ket{\psi_2}$ is much larger in extent (and in magnitude) than $\ket{\psi_0}$, and the extra node in the radial dependence due to the contribution of $2s$ is visible at about 5nm. Similarly, the state $\ket{\psi_3}$ is much larger in extent (and in magnitude) than $\ket{\psi_1}$.

The square bracket in Eqn (\ref{chiN}) gives the scaling of $\chi^{(N)}$ from hydrogenic atoms in vacuum to hydrogenic impurities in semiconductors, just as that in Eqn (\ref{rate}) does for $w^{(N)}$, and as before, the much smaller $I_a$ greatly increases the strength of the non-linearity. For example, the low-frequency limit of  the hyperpolarizability  $\chi^{(3)}/n_{\text{3D}}$ for Si:P is much larger than that for hydrogen or alkali metal vapours such as Rb \cite{MH}, as shown in Fig~\ref{fig:chi3spectrum}. 

Some of the highest values of $\chi^{(3)}$ have been reported for solids, e.g., 
$2.8\times 10^{-15}$ $\text{m}^2/\text{V}^2$ for InSb
\cite{Yuen1982}
and  $2 \times 10^{-16}$ $\text{m}^2/\text{V}^2$ for GaTe
\cite{Susoma2016}. 
To convert the hyperpolarizability to a bulk $\chi^{(3)}$ value requires the concentration. To match InSb with Si:P at low frequency where $C^{(3)}\approx 1$ (Fig~\ref{fig:Ucceffect}) (and $\chi^{(3)}/n_{\text{3D}}=2.9 \times 10^{-38}$ $\text{m}^5/\text{V}^2$) requires a donor density of $n_{\text{3D}}= 10^{17}$ $\text{cm}^{-3}$ (where the donor-donor distance is 10$a_B$). At high frequency, the hyperpolarizability is much higher, but the density should be lower to avoid inhomogeneous concentration broadening of the nearby excited levels. For example, $C^{(3)}\approx 20$ between the $2p_0$ and $2p_\pm$ resonances at $\omega=\bar{\omega}_{2p}/3= 2\pi \times 3.2$ THz  (Fig~\ref{fig:Ucceffect}), and we match InSb at a density of $n_{\text{3D}}= 5 \times 10^{15}$ $\text{cm}^{-3}$ at which concentration the 2p lines are well resolved \cite{Thomas1981}. If $3\omega$ is moved even closer to the $2p_\pm$ resonance (or if the resonance is tuned with a magnetic field \cite{murdin2013}), then $\chi^{(3)}$ could easily exceed InSb. Losses due to dephasing by phonon scattering may become important if the time spent in the intermediate states exceeds the phonon lifetime. Since the inverse of the former is given approximately by  the detuning ($\Delta f \Delta t \ge 1/2\pi $) and the inverse phonon-limited width ($1/\pi T_2=1$ GHz \cite{Steger2009,Greenland2010}), then this loss is negligible for much of the spectrum. At 50 GHz below the $2p_\pm$ line so that such losses may be ignored, $C^{(3)}\approx 200$, and $\chi^{(3)}$ is an order of magnitude above InSb. 

We are not aware of any larger values for bulk media, but  higher ``bulk" values have been reported for 2D systems such as graphene and MoS$_2$ for which  $\chi^{(3)}L$ data are divided by an interaction thickness $L$ to obtain $\chi^{(3)}$; in particular, reports for graphene range from  $10^{-19}$ \cite{Woodward2017,Karvonen2017} to  $10^{-15}$ $\text{m}^2/\text{V}^2$
\cite{Saynatjoki2013} for near-IR excitation and up to $10^{-10}$ $\text{m}^2/\text{V}^2$ in the THz region under resonant enhancement by landau levels in a magnetic field \cite{KonigOtto2017}. In the case of coupled quantum wells, large values of $ \chi^{(3)} $ may be engineered through resonances, as demonstrated up to $10^{-14}$ $\rm{m}^2/\rm{V}^2$  \cite{Sirtori}. However, since the non-linear effect is limited by the interaction length,  the 2D $\chi^{(3)}L$ is probably a better figure of merit in these cases, and for THz field-enhanced graphene with 50 layers, $\chi^{(3)}L=9\times 10^{-20}\text{m}^3/\text{V}^2$ \cite{KonigOtto2017}, or $\chi^{(3)}L=1.4\times 10^{-18}$ $\text{m}^3/\text{V}^2$ for resonant coupled QWs \cite{Sirtori}. Even higher values are predicted for doped QWs up to $\chi^{(3)}L=5\times 10^{-17}$ $\text{m}^3/\text{V}^2$ \cite{Yildirim2011}. To match this value with Si:P at $\omega=\bar{\omega}_{2p}/3=2\pi \times 3.2$ THz and $n_{\text{3D}}= 5 \times 10^{15}$ $\text{cm}^{-3}$ (see above) would require a sample thickness of $L=2$ cm. Obviously, the required thickness can be significantly reduced when close to resonance, or for germanium. 

\subsection{Efficient third-harmonic generation}
The non-linear susceptibility is important for predicting the strength of frequency conversion processes such as third-harmonic generation (3HG), and we use this as an example application to investigate the utility of the medium. A solution for the amplitude of the generated wave produced by 3HG, neglecting absorption, is given by \cite{Shen}. Converting to irradiance in MKS units,
\begin{equation}
   \frac{I_{\text{out}}}{I_{\text{in}}} = \left( \frac{3\omega_{\text{in}} \chi^{(3)}L I_{\text{in}}}{4  \epsilon_0 n^2 c^2}  \right) ^2 =  \left( \frac{   I_{\text{in}} f_{\text{in}} n_{\text{2D}} }{x} C^{(3)} \right) ^2
\end{equation}
where $I_{\text{in}}$ is the irradiance of the input pump wave at frequency $f_{\text{in}}$, and $n$ is the geometric mean of the refractive indexes for the input and output waves, and $n_{\text{2D}}=n_{\text{3D}}L$. Note that the isotropy mentioned earlier means that the polarization of the input and output waves must be parallel.  We ignored a factor for the phase matching, which is unity if the length of the sample $L \ll L_c$, where the coherence length $L_c=\pi c/(3\omega_{\text{in}}[n_{\text{out}}-n_{\text{in}}])$. Si:P at room temperature has a nearly constant $n=3.4153$ in the range from 1 THz to 12 THz \cite{Chick}, leading to typical values of $L_c\approx 10$ cm. The factor $x= 6.9 \times 10^{23} \text{W/cm}^2\times \text{THz} \times \text{cm}^{-2} $ for silicon. For comparison, germanium has $x= 9.2 \times 10^{19} \text{W/cm}^2\times \text{THz} \times \text{cm}^{-2} $.

To illustrate the possible applications of this high $\chi^{(N)}$, we note that two types of THz diode lasers are available, the quantum cascade laser (QCL)   from 0.74 THz \cite{Scalari2010} to 5.4 THz \cite{Wienold2015} with output powers of up to a few W \cite{Li2014,Li2017}, and the hot hole (p-Ge) laser \cite{Pfeffer1993,hubers2005} with a similar range and power. However, there is a large gap in the availability of solid-state sources  from about 5 THz to about 12 THz \cite{Ohtani2014}, where the GaAs Reststrahlen band renders laser operation impossible. This is an important region for quantum qubit applications \cite{Saeedi2015,Litvinenko2015,Chick2017,Stoneham2003}. Currently, the gap is only filled by larger, more expensive systems (difference frequency generators and free electron lasers). Tripling the output of 2-4 THz QCLs would fill the gap, but their output powers are far smaller than those typical for a pump laser in standard tripling applications, so a giant non-linearity is critical. At $\omega=\bar{\omega}_{2p}/3=2\pi \times 3.2$ THz, $C^{(3)}\approx 20$, so for $n_{\text{2D}}= 10^{16}$ $\text{cm}^{-2}$ (see above), a 1\% predicted conversion may be obtained with 100 kW/cm$^2$, and by moving to 50 GHz below the $\text{2p}_{\pm}$ resonance this value could be brought down to 10 kW/cm$^2$, which is just about achievable with a well focussed QCL, and would thus provide enough output for spectroscopy applications. 
A nonlinear process that may possibly reduce the 3HG efficiency is multiphoton ionization \cite{bebb1966} since it reduces the population of the donors in the ground state. When $\omega=\bar{\omega}_{2p}/3$, for example, a four-photon absorption takes the electron to the continuum. We estimate this ionization in Si:P using the implicit summation method and find the rate is $w=3.17$ $\text{s}^{-1}$ for $I_{\text{in}}=10$ $\text{kW/cm}^2$. This result simply means that the pulses must be kept significantly shorter than a second to avoid significant ionization.

In summary, we calculated the absolute values of the THz non-linear coefficients for the most common semiconductor materials, lightly doped silicon and germanium, which are available in the largest, purest and most regular single crystals known. The values we obtain for off-resonance rival the highest values obtained in any other material even when resonantly enhanced, and the material could gain new applications in THz photonics. We also predict the highly efficient third-harmonic generation of THz light in doped silicon and germanium. Our multi-valley theory for nonlinear optical processes of donors in silicon and germanium can be readily applied to any donor in any semiconductor host in which the effective mass approximation is reliable.

\section{Materials and methods}
Details of the finite element computation used for solving the coupled partial differential equations (Eqns~\eqref{eq:coupledPDEs}) are provided in the Supplementary Material.

\textbf{Data availability}: Data for Nguyen Le et al. “Giant non-linear susceptibilityof hydrogenic donors in silicon and germanium", https://doi.org/10.5281/zenodo.1257357. The data underlying this work is available without restriction.

\section{Acknowledgements}
We acknowledge financial support from the UK Engineering and Physical Sciences Research Council [ADDRFSS, Grant No. EP/M009564/1] and EPSRC strategic equipment grant no. EP/L02263X/1.

\section{Conflict of interests}
None of the authors has any conflict of interest.

\section{Contributions}
N.H. Le and B.N. Murdin worked on the multivalley theory and the finite element calculation of the third-order susceptibility. B.N. Murdin and G.V. Lanskii calculated the third-harmonic generation efficiency. B.N. Murdin, N.H. Le and G. Aeppli wrote the manuscript. All authors contributed to the discussion of the results.

\end{document}

%% ****** Title of paper ****** %
\title{Supplementary materials for Nguyen Le et al ``Giant non-linear susceptibility in silicon hydrogenic donors''}

\author{Nguyen H. Le}
\affiliation{ Advanced Technology Institute and SEPNet, University of Surrey, Guildford, GU2 7XH, United Kingdom}
\author{Grigory V. Lanskii}
\affiliation{Institute of Monitoring of Climatic and Ecological Systems SB RAS, 10/3 Academical Ave., Tomsk 634055, Russia}                                  
\author{Gabriel Aeppli}
\affiliation{ Laboratory for Solid State Physics, ETH Zurich, Zurich, CH-8093,Switzerland}
\affiliation{  Institut de Physique, EPF Lausanne, Lausanne, CH-1015, Switzerland} 
\affiliation{ Swiss Light Source, Paul Scherrer Institut, Villigen PSI, CH-5232, Switzerland}
\author{Benedict N. Murdin}
\affiliation{ Advanced Technology Institute and SEPNet, University of Surrey, Guildford, GU2 7XH, United Kingdom}

\maketitle

There are several theoretical innovations described in the main paper, and in this supplementary materials we provide the derivations. Implicit summation has already been developed for hydrogen, and several versions of the multivalley effective mass for silicon donors are available with different techniques and levels of approximation. Here we have developed:
\begin{enumerate}
\item The implicit summation technique for  multiphoton non-linear susceptibility and multiphoton absorbtion in a multivalley donor, including an approximate central cell correction.

\item A successful but simple 2D Finite Element Method (FEM) for solving for the partial differential equations that appear in the implicit summation technique, including the inter-valley coupling due to the CCC, that can be written in a few lines of code in standard packages (we used Mathematica). The same method can be used to obtain the wavefunction and energy of a multivalley donor. 

\end{enumerate}

In order to implement the implicit summation technique for the multiphoton process we will require solution of  the Schr\"{o}dinger equation $H\ket{\psi_j}=\hbar\omega_j\ket{\psi_j}$ for the ground state and for the excited state in the case of absorption, and we will also require solution of PDEs of the form $[H-W_n]\ket{\psi_n}=\epsilon.\mathbf{r}\ket{\psi_{n-1}}$ where $H$ is the Hamiltonian and $\ket{\psi_0}$ is the ground state. We therefore require a prescription for both diagonalising $H$ and using it in a PDE, ideally consistently. In the main text we derive the Schrodinger equation and the PDEs for the intermediate states in the valley-specific spherical coordinates of the lab frame, in practice it is better to work in a stretched frame where the kinetic term is isotropic as the FEM converges faster with increasing mesh density in this frame (see below). In this Supplementary Materials all the equations are expressed in the spherical coordinates of this stretched frame. 

\section{Multivalley Schr\"{o}dinger Equation}

Silicon is indirect with six equivalent conduction band minima (valleys) near the Brillouin zone edge X-points, each characterized by Fermi surfaces that are prolate ellipsoids, characterized by the effective mass ratio $\gamma$. According to the Kohn-Luttinger effective mass approximation (EMA) \cite{KL1955}, the wavefunction $\psi_{j}(\bm{r})$ of a shallow donor can be decomposed into six equivalent, slowly varying hydrogenic envelope functions, each modulated by plane wave functions corresponding to crystal momentum at the minima $\mathbf{k}_{\mu}$ (and a lattice periodic function that is unimportant here). The Hamiltonian comprises a kinetic energy term that operates only on the envelope function, the slowly varying Coulomb potential of the donor $ V (\mathbf{r}) $, and a rapidly varying potential due to the quantum defect that is short range, $ U ( \mathbf{r} )  $, referred to as the central cell correction (CCC). The kinetic energy has cylindrical symmetry because $\gamma \neq 1$, and $V( \mathbf{r} ) \& U( \mathbf{r} )$ are spherical and tetrahedral  respectively. The first two terms produce wavefunctions that are approximately hydrogen-like and the third mixes them to produce states that transform as the $A_1, E$ and $T_2$ components of the $T_d$ point group. We write 
\begin{equation}\label{multivalley_state}
 \psi_j(\bm{r}) = \sum_{\mu=\pm1,\pm2,\pm3}  e^{  i\mathbf{k}_\mu .\mathbf{r} } u_{\bm{k}_{\mu}}(\bm{r})  F_{j,\mu}(\bm{r}),
\end{equation}
where $F_{j,\mu}(\bm{r})$ is the envelope function and the valley index $\mu$ runs over $\pm 1,\pm 2, \pm 3$ where 1,2,3 are the three crystal axes. For simplicity of derivation we first assume that $u_{\bm{k}_{\mu}}(\bm{r})=1$. A more rigorous derivation where this assumption is relaxed is provided in Sec. III. We note that there is no change in the form of the final equations.

Substituting Eqn~\eqref{multivalley_state} (with $u_{\bm{k}_{\mu}}(\bm{r})=1$)  into the Schr\"{o}dinger equation
\begin{equation}
H\ket{\psi_j} = \hbar\omega_j\ket{\psi_j} 
\end{equation} 
within the EMA produces
\begin{equation}\label{EMA}
  \sum _{\mu=\pm1,\pm2,\pm3} e^{  i\mathbf{k}_\mu .\mathbf{r} }  \left[ H_0+U(\bm{r}) -\hbar\omega_j\right] F_{j,\mu}(\bm{r}) =0,
\end{equation}
where 
\begin{align}\label{eq:H0xyz}
H_0=-\frac{E_H a_B^2}{2}\left[\frac{\partial^2}{\partial x^2}+\frac{\partial^2}{\partial y^2}+\gamma \frac{\partial^2}{\partial z^2}\right]-\frac{E_H a_B}{\sqrt{x^2+y^2+z^2}},
\end{align}
and the Coulomb potential $V(\bm{r})$ is the last term. Here $\gamma=m_t/m_l$, $a_B=4\pi\epsilon_0\epsilon_r\hbar^2/m_t e^2$ and $E_H=m_t e^4/(4\pi\epsilon_0\epsilon_r)^2$ where $m_t$ is the transverse effective mass, $m_l$ the longitudinal effective mass and $\epsilon_r$ the dielectric constant of silicon. It has been written using a valley-specific coordinate system $x,y,z$ where $z$ is the valley axis, i.e. this frame is rotated relative to the lab frame $x_1,x_2,x_3$. In all this work $x,y,z$ (and later $r',\theta',\phi'$) refer to the valley coordinate frame and $x_1,x_2,x_3$ refer to the lab frame crystal axes.

The plane wave factors are quickly varying compared with every other factor except  $U(\mathbf{r})$. Since $U(\bm{r})$ is very short-range, it effectively samples the wavefunction amplitude at the origin.
Premultiplying Eqn~\eqref{EMA} by  $ e^{  -i\mathbf{k}_{\mu'}.\mathbf{r} } $  and averaging over a volume  $  \left( 2 \pi /k_0 \right) ^{3} $ around $\mathbf{r}$ where $k_0=\vert \mathbf{k}_{ \mu }\vert$  produces
\begin{equation}\label{eq:mvscr}
  \left[ H_0 -\hbar \omega_j\right]F_{j,\mu}(\bm{r}) +\sum _{\mu'=\pm1,\pm2,\pm3} U_{ \mu \mu'} \delta(\mathbf{r}) F_{j,\mu'}(\bm{r}) =0,
\end{equation}
where  $\delta(\bm{r})$ is the Dirac delta function and 
\begin{align}\label{Uccdef}
U_{\mu \mu'}=\int d\bm{r}\, e^{i (\bm{k}_{\mu'}-\bm{k}_{\mu}).\bm{r}}U(\bm{r}).
\end{align} 
Clearly the $U$ term only affects states with significant amplitude at $\mathbf{r}=0$ due to the $\delta$-function, i.e. principally the $1s$ state. 

For an $A_{1}$ state all the envelope functions have the same amplitude at $r=0$, hence  $\sum _{\mu'} U_{\mu\mu'}  \delta(\bm{r}) F_{j,\mu'}(\bm{r})=-U_{cc} \delta(\bm{r})  F_{j,\mu}(\bm{r})$ where $U_{cc}=- \sum _{\mu'} U_{\mu\mu'}$. It is found experimentally that for $E$ and $T_2$ states the CCC has rather small effect and so we neglect it. The eigenproblems for $A_1$ and for $E,T_2$ states are then respectively:
\begin{eqnarray}
 \left[ H_0  - U_{cc} \delta({\mathbf{r}})  -\hbar \omega _j \right] F_{j, \mu}(\bm{r})  =0, \label{eigenproblemA1}\\
 \left[ H_0 -\hbar \omega _{j} \right] F_{j,\mu}(\bm{r})  =0 \label{eigenproblemE}. 
\end{eqnarray}
These equations are the same for all $\mu$, and we solve them using a valley-specific coordinate system $x,y,z$ where $z$ is the valley axis. The states with odd parity envelopes have  $ F_{j,\mu}(0) =0 $, so the $A_1$, $E$ and $T_2$ states are degenerate, whereas even parity envelopes have  $ F_{j,\mu}(0)  \neq 0 $  and the central cell correction lowers the energy of the $A_1$ states.

\section{Finite element method for hydrogenic donors with anisotropic mass and central cell correction}

Equations \eqref{eigenproblemA1} and \eqref{eigenproblemE} can be solved with a Finite Element Method (FEM). We first perform a sequence of transformations in order to achieve a combination of simplification of the problem and  improving the rate of convergence as the FEM mesh density is increased. First we transform $x,y,z$ to an anisotropic frame that symmetrizes the kinetic energy (with the side-effect that the Coulomb potential is made anisotropic):
\begin{eqnarray}\label{eq:strframe}
x&=&x',\nonumber \\
y&=&y', \nonumber \\
z&=&\sqrt{\gamma} z',
\end{eqnarray}
(and obviously $z'$ is still the valley axis, and $x',y',z'$ are still rotated relative to the lab frame) producing
\begin{equation}
H_0^s=-\frac{E_Ha_B^2}{2}\left[\frac{\partial^2}{\partial x'^2}+\frac{\partial^2}{\partial y'^2}+ \frac{\partial^2}{\partial z'^2}\right]-\frac{E_Ha_B}{\sqrt{x'^2+y'^2+\gamma z'^2}}.
\end{equation}
We now transform to spherical polar coordinates to take advantage of the spherical symmetry of the kinetic energy term, and again because it improves the FEM convergence
\begin{eqnarray}
x'&=&r'\sin\theta'\cos\phi',\nonumber \\
y'&=&r'\sin\theta'\sin\phi', \nonumber \\
z'&=&r'\cos\theta',
\end{eqnarray}
where $\theta'$ is the zenith angle away from $z'$ and $\phi'$ is the azimuthal angle around $z'$ away from $x'$, and the symmetrized polar Hamiltonian is
\begin{equation}
H_0^{sp}=-\frac{E_Ha_B^2}{2}\left[
\frac{1}{r'^2}\frac{\partial}{\partial r'}\left( r'^2 \frac{\partial}{\partial r'}\right) +\frac{1}{r'^2\sin^2\theta'}\frac{\partial^2}{\partial \phi'^2}+ \frac{1}{r'^2\sin\theta'}\frac{\partial}{\partial \theta'}\left(\sin\theta' \frac{\partial}{\partial \theta'}\right)
\right]
-\frac{E_Ha_B}{r'\sqrt{1-(1-\gamma) \cos^2\theta'}}.
\end{equation}
We choose the wavefunction 
\begin{equation}\label{wavefn}
F_{j,\mu}(\bm{r})=e^{im \phi'}f_{j,m,\mu}(r',\theta') \equiv e^{im \phi'} \frac{1}{r'} Y_{j,m}(r',\theta'),
\end{equation}
and there is no $\mu$ dependence on the RHS because all valleys are equivalent and the coordinates are valley-oriented. Since $m$ is a good quantum number we now have a 2D problem and Eqns~\eqref{eigenproblemA1} and \eqref{eigenproblemE} become
\begin{eqnarray}
 \left[ H_0^{sp2d}  - (U_{cc}/\sqrt{\gamma})  \delta(\bm{r'})  -\hbar \omega _j \right] Y_{j,m}(r',\theta')  =0 \nonumber \\
 \left[ {H}_0^{sp2d} -\hbar \omega _{j} \right] Y_{j,m}(r',\theta')  =0. 
\end{eqnarray}
where the symmetrized polar 2D Hamiltonian is
\begin{equation}\label{eq:H0sp2d}
H_0^{sp2d}=-\frac{E_Ha_B^2}{2}\left[
\frac{\partial^2}{\partial r'^2} -\frac{m^2}{r'^2\sin^2\theta'}+ \frac{1}{r'^2\sin\theta'}\frac{\partial}{\partial \theta'}\left(\sin\theta' \frac{\partial}{\partial \theta'}\right)
\right]
-\frac{E_Ha_B}{r'\sqrt{1-(1-\gamma) \cos^2\theta'}}.
\end{equation}
Finally, we compress the radial scale to a tangent space, which transforms the semi-infinite $r\theta$ plane onto the finite rectangle $ 0<\eta< \pi/2 ,0 <\theta< \pi $, and again we find that it gives the improved convergence with increasing FEM mesh density: 
\begin{eqnarray}
r'&=&r_0\tan\eta,
\end{eqnarray}
The constant $r_0$ is a scaling factor, and numerical experiments show that it should be chosen to be comparable to the radius of wavefunction of interest for an accurate result.
Now we have 
\begin{eqnarray}
 \left[ H_0^{sp2dc}  - (U_{cc}/\sqrt{\gamma})  \delta(\bm{r'})   -\hbar \omega _j \right] y_{j,m}(\eta,\theta')  =0, \label{eigenproblemUcc} \\
 \left[ {H}_0^{sp2dc} -\hbar \omega _{j} \right] y_{j,m}(\eta,\theta')  =0, \label{eigenproblem}
\end{eqnarray}
where the symmetrized polar 2D compressed Hamiltonian is
\begin{equation}\label{eq:H0tan}
{H}_0^{sp2dc}=-\frac{E_Ha_B^2}{2r_0^2}\left(
\cos^2\eta\frac{\partial}{\partial \eta}\left(\cos^2\eta\frac{\partial}{\partial \eta} \right) -\frac{m^2\cot^2\eta}{\sin^2\theta'}+ \frac{\cot^2\eta}{\sin\theta'}\frac{\partial}{\partial \theta'}\left(\sin\theta' \frac{\partial}{\partial \theta'}\right)
\right)
-\frac{E_Ha_Br_0^{-1}\cot\eta}{\sqrt{1-(1-\gamma) \cos^2\theta'}}.
\end{equation}
For our numerical calculation we model the Dirac delta function $\delta(\bm{r'})$ by a short range step function $\Theta (r_{cc}-r')/(4\pi r_{cc}^3/3)=\Theta (r_{cc}-r_0 \tan \eta)/(4\pi r_{cc}^3/3)$ where $r_{cc}\ll a_B$. We fix $r_{cc}=0.1 a_B$ and use a bisection procedure to find the value of $U_{cc}$ that reproduces the experimental energy of the $1\text{sA}_1$ state in Eqn~\eqref{eigenproblemA1}.

The net transformation is
\begin{eqnarray}
x&=&r_0\tan\eta\sin\theta'\cos\phi',\nonumber \\
y&=&r_0\tan\eta\sin\theta'\sin\phi', \nonumber \\
z&=&\sqrt{\gamma} r_0 \tan\eta\cos\theta'.
\end{eqnarray}
The wavefunctions should be normalized with
\begin{equation}
Z= 2\pi r_0 \sqrt{\gamma} \int_0^{\pi/2} d \eta \sec^2\eta \int_0^\pi d\theta'    \sin\theta' \vert y_{j,m}(\eta,\theta') \vert ^2.
\end{equation}

We can classify the wavefunction $y_{j,m}(\eta,\theta')$ by parity $p$ and the quantum number $m$, with $m$ an integer. The functions with $p,m$ both odd or both even satisfy the symmetric condition $y(\eta,\theta')=y(\eta,\pi-\theta')$, while the functions with $p,m$ opposite parity satisfy the anti-symmetric condition $y(\eta,\theta')=-y(\eta,\pi-\theta')$. Therefore we need to solve for $y(r,\theta')$ only in the domain $0\leq \theta' \leq \pi/2$. The boundary conditions for symmetric and anti-symmetric functions in $\theta'$ are
\begin{align}
\Gamma_s: \ &y(0,\theta')=y(\pi/2,\theta')=0, \nonumber \\
\Gamma_a:\ &y(0,\theta')=y(\pi/2,\theta')=y(\eta,\pi/2)=0.
\end{align}
In our numerical calculation we choose $\eta_{\text{max}}=\pi/2.1$ to avoid divergence in the numerical integral of the multiphoton matrix elements.

\section{Intervalley coupled multiphoton PDEs}\label{sec:mvPDE}
\subsection{Silicon}
\begin{figure}
\centering
\includegraphics[scale=0.5]{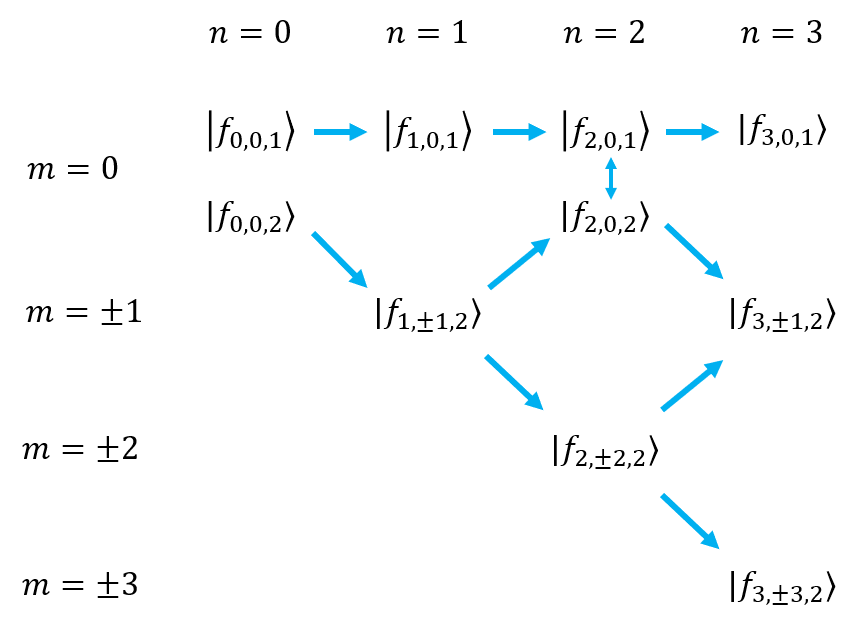}
\caption{\label{fig:interactions} Multiphoton intermediate states $  \vert f_{n,m,\mu} \rangle $ and their interactions produced by  dipole excitation along $x_1$ (horizontal arrows for the $x_1$ valley and diagonal arrows for the $x_2$ valleys) and produced by $U_{cc}$ (vertical arrows). NB the $x_3$ valley function is the same as for the $x_2$ valley. For calculation of $\chi^{(N)}$ only  $  \vert f_{N,0,1} \rangle $ and $  \vert f_{N,\pm 1,2} \rangle $ are  needed.}
\end{figure}

In this section we derive the intervalley coupled PDEs that appear in the implicit summation technique (Eqn M8 in the main text). After obtaining the ground state $\ket{\psi_0}=\ket{\psi_g}$ we want to solve the PDE
\begin{equation}\label{multiphotonPDE}
(H-W_n)\ket{\psi_n}=E_H \zeta\ket{\psi_{n-1}},
\end{equation}
where $\zeta=\bm{\epsilon}.\bm{r}/a_B$, and we use a similar expansion for the multivalley intermediate state wavefunctions as was used for the eigenstates of $H$ in Sec. I.
\begin{equation}\label{eq:n_expansion_mu}
 \psi_n(\bm{r}) = \sum _\mu  e^{  i\mathbf{k}_{ \mu }.\mathbf{r} } u_{\bm{k}_{\mu}}(\bm{r}) F_{n,\mu}(\bm{r}), 
 \end{equation}
and $\psi_0(\bm{r})$ is given by the $A_1$ ground state, hence $F_{0,\mu}(\bm{r})$ can be obtained from solving Eqn~\eqref{eigenproblemA1}. Again assuming $u_{\bm{k}_{\mu}}(\bm{r})=1$ as in Sec. I and substituting into Eqn ~\eqref{multiphotonPDE}, we obtain within the EMA
\begin{equation}
  (H-W_n)\psi_n(\bm{r}) = \sum _\mu  e^{  i\mathbf{k}_{ \mu }.\mathbf{r} } [H_0+U(\bm{r})-W_n] F_{n,\mu }(\bm{r}) =E_H\zeta\sum _\mu  e^{  i\mathbf{k}_{ \mu }.\mathbf{r} }  F_{n-1,\mu}(\bm{r}). 
\end{equation}
Premultiplying by 
$ e^{  -i\mathbf{k}_{ \mu ' }.\mathbf{r} } $  and averaging over a volume  $  \left( 2 \pi /k_0 \right) ^{3} $ around $\mathbf{r}$  produces
 \begin{align}\label{eq:mvpde}
 \left[ H_0-W_n\right]F_{n,\mu}(\bm{r}) +\sum _{\mu'}  U_{\mu\mu'} \delta(\mathbf{r}) F_{n,\mu'}(\bm{r}) = E_H \zeta F_{n-1,\mu}(\bm{r}), 
\end{align}
where  $U_{\mu\mu'}$ is given in Eqn \eqref{Uccdef}. In this paper we neglect the energy shift due to the CCC of the $1s\text{E}$ and $1s\text{T}_2$ states, and this is equivalent to assuming that $U_{\mu\mu'}$ is nearly the same for all $\mu, \mu'$. Let us denote  $U_\parallel$ for  $\mu'=\mu$, $U_\pm$ for $\mu'=-\mu$, and $U_\perp$ for all other combinations, in a perturbative treatment of the CCC the energy shift of the $1s \text{A}_1$,  $1s \text{E}$ and $1s \text{T}_2$ are given by $(U_\parallel+U_\pm+4U_\perp)|F_{\text{KL}}(0)|^2$, $(U_\parallel+U_\pm-2U_\perp)|F_{\text{KL}}(0)|^2$, and $(U_\parallel-U_\pm)|F_{\text{KL}}|(0)|^2$, respectively \cite{saraiva2015}, where $F_{\text{KL}}(\bm{r})$ is the unperturbed single-valley envelope function of the ground state without the CCC, as discussed by Kohn and Luttinger \cite{KL1955}. Neglecting the shift in the $1s \text{E}$ and $1s \text{T}_2$ states means that
\begin{align}
U_\parallel \approx U_\pm \approx U_\perp=-U_{cc}/6,
\end{align}
so we can replace $U_{\mu\mu'}$ by $-U_{cc}/6$. By symmetry (i.e knowing that we start from the symmetric $A_1$ ground state which has $\ket{g,\mu}$ the same for each valley with equal amplitudes, and knowing that the  polarization $\zeta$ is the only factor that breaks the symmetry), $F_{n,\mu}=F_{n,-\mu}$, so each $U_{cc}$ term is doubled up and there are three coupled equations:
\begin{eqnarray}  
\left[H_0-W_n-\frac{U_{cc}}{3}\delta(\mathbf{r})\right]F_{n,1}(\bm{r}) - \frac{U_{cc}}{3}\delta(\mathbf{r}) \left[F_{n,2}(\bm{r})+F_{n,3}(\bm{r}) \right]&=& E_H \zeta F_{n-1,1}(\bm{r}),\nonumber \\
  \left[ H_0-W_n-\frac{U_{cc}}{3}\delta(\mathbf{r})\right]F_{n,2}(\bm{r})-\frac{U_{cc}}{3}\delta({\mathbf{r}})\left[F_{n,1}(\bm{r})+F_{n,3}(\bm{r})\right] &=& E_H\zeta F_{n-1,2}(\bm{r}), \nonumber \\
\left[ H_0-W_n-\frac{U_{cc}}{3}\delta({\mathbf{r}})\right]F_{n,3}(\bm{r})-\frac{U_{cc}}{3}\delta({\mathbf{r}})\left[F_{n,1}(\bm{r})+F_{n,2}(\bm{r})\right] &=& E_H\zeta F_{n-1,3}(\bm{r}).
\end{eqnarray} 
In the above equations the Hamiltonian $H_0$ is given by Eqn~\eqref{eq:H0xyz} with the valley-specific coordinates in $H_0$ understood to be belong to the valley of the envelope function that it acts on. We wish to transform to the same frame as used for the eigenstate calculation. In solving for the eigenstates we use wavefunctions like Eqn~\eqref{wavefn} in the valley-specific anisotropic spherical polar frame $(r',\theta',\phi')$ so that $\theta' \& \phi'$ are relative to the valley axis: 
\begin{equation}\label{eq:n_expansion_m}
 F_{n,\mu}(\bm{r})=\sum _{ m }e^{im \phi'} f_{n,m,\mu}(r',\theta')\equiv \sum _{ m }\frac{1}{r'} e^{im \phi'}  Y_{n,m,\mu}(r',\theta'), 
\end{equation}
and substitute into the PDEs, premultiply by $e^{im'\phi'}$ and integrate over $\phi'$. The operator $\zeta$ is polarization dependent, and the valley specific coordinates make it valley-specific too. In the case of [100] polarized light i.e. $a_B \zeta=x_1$, we have $a_B \zeta_1=z=\sqrt{\gamma}r'\cos\theta',a_B \zeta_2=x=r'\sin\theta'\cos\phi', a_B \zeta_3=y=r'\sin\theta'\sin\phi'$, and we obtain
\begin{eqnarray}\label{eq:mvpdeY}  
\left[ H_0-W_n-\mathcal{D} \right]Y_{n,m,1} - \mathcal{D} (Y_{n,m,2}+ Y_{n,m,3}) &=& (E_H/a_B)  \sqrt{\gamma}r'\cos\theta' Y_{n-1,m,1}, \nonumber \\
  \left[ {H}_0-W_n-\mathcal{D}\right]Y_{n,m,2}-\mathcal{D}(Y_{n,m,1}+Y_{n,m,3}) &=& (E_H/a_B)  r'\sin\theta' \frac{1}{2}\left(Y_{n-1,m-1,2}+Y_{n-1,m+1,2}\right), \nonumber \\
  \left[ {H}_0-W_n-\mathcal{D}\right]Y_{n,m,3}-\mathcal{D}(Y_{n,m,1}+Y_{n,m,2}) &=& (E_H/a_B)  r'\sin\theta' \frac{1}{2i}\left(Y_{n-1,m-1,3}- Y_{n-1,m+1,3}\right),
\end{eqnarray} 
where $H_0$ is given in Eqn~\eqref{eq:H0sp2d}, $\mathcal{D}=(U_{cc}/3\sqrt\gamma)\delta(\bm{r}')\delta_{m,0}$ and $\delta_{m,0}$ is the Kronecker Delta. In the above equations we drop the valley-specific coordinates in $Y_{n,m,\mu}$ for notational simplicity, and again the coordinates in $H_0$ and $\zeta$ belong to the valley of the envelope function that they act on. Note that the off-diagonal terms on the LHS are zero unless $m=0$. The second and third equations are equivalent with 
\begin{equation}
Y_{n,m,3}=i^{-m}Y_{n,m,2},
\end{equation}
so in the tangent space the $\Upsilon_{n,m,\mu}(\eta,\theta')$ are
\begin{eqnarray}  
\left[ {H}_0-W_n-\mathcal{D} \right]\Upsilon_{n,m,1} - 2 \mathcal{D} \, \Upsilon_{n,m,2} &=& (E_H/a_B) \sqrt{\gamma}r_0 \tan \eta\cos \theta' \, \Upsilon_{n-1,m,1}, \nonumber \\
 \left[ {H}_0-W_n-2\mathcal{D}\right] \Upsilon_{n,m,2}+ -\mathcal{D}\, \Upsilon_{n,m,1} &=& (E_H/a_B)  r_0 \tan\eta\sin\theta' \frac{1}{2}\left(\Upsilon_{n-1,m-1,2}+\Upsilon_{n-1,m+1,2}\right), \nonumber 
\end{eqnarray} 
where the Hamiltonian is now given by Eqn~\eqref{eq:H0tan}. It is evident that these equations are coupled by    $ U_{cc} $  when  $ m=0 $ and the parity is even (shown by the vertical arrows on  Fig.~\ref{fig:interactions}).

In the case of an arbitrary polarization $\bm{\epsilon}$, since we are free to rotate the valley-specific coordinates around the valley axis, we can always choose a coordinate such that $\bm{\epsilon}$ has only the components $\epsilon^{\parallel}_{\mu}$ along the $z$ axis and $\epsilon^{\perp}_{\mu}$ along the $x$ axis, and the component along $y$ is zero, for \emph{all} valleys $\mu$. Repeating the steps for deriving Eqn \eqref{eq:mvpdeY} we obtain
\begin{eqnarray}\label{eq:generalpol}
 \left[{H}_0-W_n \right] Y_{n,m,\mu} - \mathcal{D} \sum_{\mu'} Y_{n,m,\mu'} &=&  \frac{E_H}{a_B}r' \left[ \epsilon^{\parallel}_{\mu}\sqrt{\gamma} \cos\theta' Y_{n-1,m,\mu}+\epsilon^{\perp}_{\mu}\frac{\sin\theta'}{2} \left(Y_{n-1,m-1,\mu} + Y_{n-1,m+1,\mu}\right)\right], 
 \end{eqnarray}
for $\mu,\mu'=1,2,3$. By utilizing this free choice of coordinate rotation we see that the envelope functions depend only on the angle between the polarization vector and the valley axes. For polarization along the [1,1,1] crystal axis, for example, $\epsilon^{\parallel}_{\mu}=1/\sqrt{3}$ and $\epsilon^{\perp}_{\mu}=\sqrt{2/3}$ for all $\mu$, and it is evident from Eqn \eqref{eq:generalpol} that $Y_{n,m,\mu}$ must be the same for all $\mu$, and Eqn \eqref{eq:generalpol} reduces to 
\begin{equation}
\left[{H}_0-W_n -3\mathcal{D}\right] Y_{n,m,\mu}  =  \frac{E_H}{a_B}r' \left[ \sqrt{\frac{\gamma}{3}} \cos\theta' Y_{n-1,m,\mu}+\sqrt{\frac{2}{3}} \frac{\sin\theta'}{2} \left(Y_{n-1,m-1,\mu} + Y_{n-1,m+1,\mu}\right)\right], 
\end{equation}
where $Y_{0,0,\mu}=Y_{g,0,\mu}$.

NB: Although the coordinate representation of vectors like $\bm{\epsilon}$ and functions like $Y_{n,m\mu}$ is coordinate dependent, scalars such as the NPA matrix element $M^{(N)}$ and the nonlinear susceptibility $C^{(N)}$ must be coordinate invariant. While one may obtain an apparent different form for $Y_{n,m,\mu}$ in a rotated frame, the final results for $M^{(N)}$ and $C^{(N)}$ have to stay the same.

\subsection{Germanium}
\begin{figure}[b]
\centering
\includegraphics[scale=0.45]{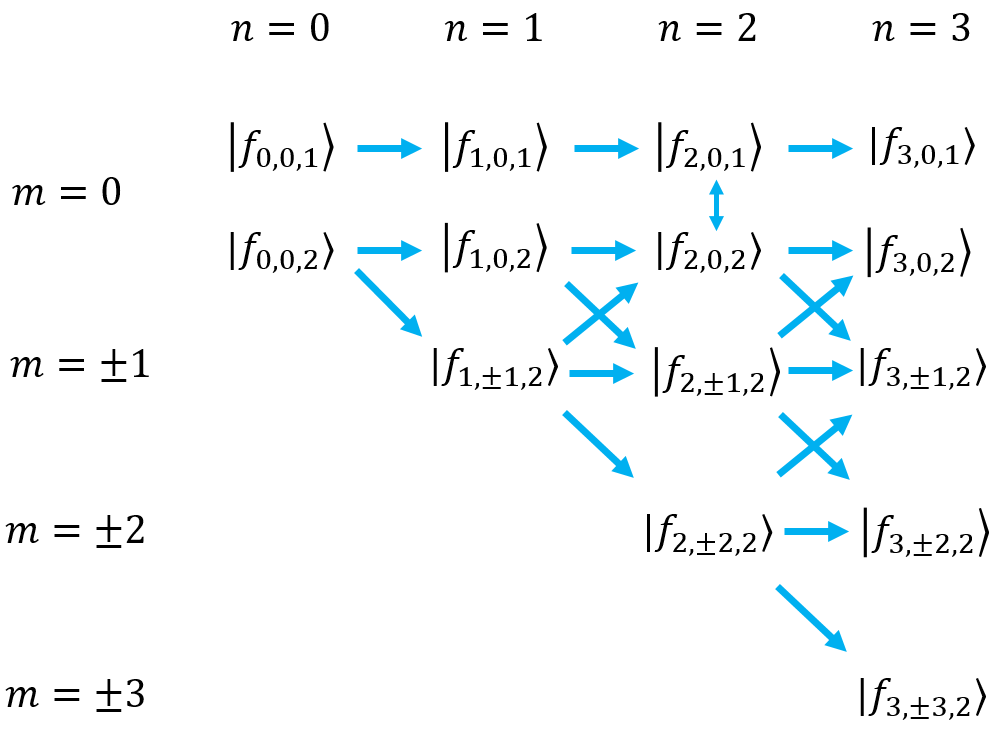}
\caption{\label{fig:intGe} Multiphoton intermediate states $  \vert f_{n,m,\mu} \rangle $ and their interactions produced by  dipole excitation (horizontal arrows and diagonal arrows) and produced by $U_{cc}$ (vertical arrows) for germanium.}
\end{figure} 

For germanium Eqn~\eqref{eq:generalpol} also applies with $\mu,\mu'=1,2,3,4$ for the four CB minima valleys along the [1,1,1] equivalent crystal axes, and $\mathcal{D} = U_{cc} \delta (\bm{r}')\delta_{m,0} /4\sqrt\gamma)$. The valley axes are given by the directional unit vectors $\bm{u}_{\mu}=[1,1,1]/\sqrt{3}$, $[-1,1,1]/\sqrt{3}$, $[1,-1,1]/\sqrt{3}$ and $[1,1,-1]/\sqrt{3}$. For light polarized along the [1,1,1] crystal axis,  the components of the polarization vector in the valley-specific coordinates are
\begin{eqnarray}
\epsilon_{1}^{\parallel}&=\bm{\epsilon}.\bm{u}_1=1, \ \ \
&\epsilon_{1}^{\perp}=0, \nonumber \\
\epsilon_{2}^{\parallel}&=\bm{\epsilon}.\bm{u}_2=1/3, \ \ \
&\epsilon_{2}^{\perp}=\sqrt{1-(\epsilon_{2}^{\parallel})^2}=2\sqrt{2}/3,  \nonumber \\
\epsilon_{3}^{\parallel}&=\bm{\epsilon}.\bm{u}_3=1/3, \ \ \
&\epsilon_{3}^{\perp}=2\sqrt{2}/3, \nonumber \\
\epsilon_{4}^{\parallel}&=\bm{\epsilon}.\bm{u}_4=1/3, \ \ \
&\epsilon_{4}^{\perp}=2\sqrt{2}/3.
\end{eqnarray}
Since $\bm{\epsilon}$ has the same components in valleys $2,3,4$, we see from Eqn~\eqref{eq:generalpol} that  $f_{n,m,2}=f_{n,m,3}=f_{n,m,4}$, and hence there are only two independent envelope functions, $f_{n,m,1}$ and $f_{n,m,2}$. Equation~\eqref{eq:generalpol} now becomes
\begin{align}
 \left[{H}_0^{(m)}-W_n-D \right] f_{n,m,1} - 3\mathcal{D} f_{n,m,2} &=  \frac{E_H}{a_B}r'  \sqrt{\gamma}\cos\theta' f_{n-1,m,1},\nonumber \\
\left[{H}_0^{(m)}-W_n-3D \right] f_{n,m,2} - \mathcal{D} f_{n,m,1} &=  \frac{E_H}{a_B}r' \left[ \frac{\sqrt{\gamma} \cos\theta'}{3} f_{n-1,m,2}+ \frac{\sqrt{2}\sin\theta'}{3}\left(f_{n-1,m-1,2} + f_{n-1,m+1,2}\right)\right],\nonumber \\
\end{align}

We used the following parameters for Ge:P: $\gamma=0.05134$, $E_H=9.40$ meV, $a_B=9.97$ nm \cite{Faulkner1969}, and $E(1s \text{A}_1)=-12.89$ meV \cite{Pajot}. The interaction between the envelope functions $f_{n,m,\mu}$ for germanium is shown in Fig.~\ref{fig:intGe}.

\section{A rigorous derivation of the intervalley coupled multiphoton PDEs}

In the previous sections we derive the intervalley coupled multiphoton PDEs by assuming $u_{\bm{k}_{\mu}}(\bm{r})=1$, which allows an intuitive and simple derivation of the equations. Here we relax this assumption. We start with the full Hamiltonian including the silicon lattice potential, and approximations within the multivalley effective mass theory are stated clearly.

The Hamiltonian for an impurity electron is $H=H_{\text{Si}}+V(r)$ where $H_{\text{Si}}=-(\hbar^2/2 m_e) \Delta + V_{\text{Si}}(r)$ and $V_{\text{Si}}(r)$ is the crystal potential of the silicon lattice. The impurity potential is $V(r)=-e^2/4\pi\epsilon_0\epsilon_r r+U(r)$, where $U(r)$ is the central cell correction. 

The first Brillouin zone of silicon can be divided into six equal subregions $\Gamma_{\mu}$,  $\mu=\pm 1,\pm 2,\pm 3$, each contains a conduction band minimum. In the one band approximation, the intermediate state $\ket{\psi_n}$ can be expanded using the Bloch states of the conduction band
\begin{align}\label{eq:blochexp}
\psi_n(\bm{r})=\sum_{\mu}\sum_{\bm{k}\in \Gamma_{\mu}}F_{n,\mu}(\bm{k})\phi_{\bm{k}}(\bm{r}),
\end{align}
where $\phi_{\bm{k}}(\bm{r})=e^{i \bm{k}.\bm{r}}u_{\bm{k}}(\bm{r})$ is the Bloch function of the pure crystal. It is the eigenstate of the unperturbed Hamiltonian
\begin{align}
H_{\text{Si}}\phi_{\bm{k}}(\bm{r})=E(\bm{k})\phi_{\bm{k}}(\bm{r}).
\end{align}
 In the approximations of the multivalley effective mass theory, we assume $F_{n,\mu}(\bm{k})$ is non-vanishing only in a small pocket of the Brillouin zone around the conduction band minimum at $\bm{k}_{\mu}$, so we can replace $u_{\bm{k}}(\bm{r})$ by  $u_{\bm{k_{\mu}}}(\bm{r})$ for each valley and Eqn~\eqref{eq:blochexp} becomes
\begin{align}
\psi_n(\bm{r})=\sum_{\mu}F_{n,\mu}(\bm{r})e^{i \bm{k_{\mu}}.\bm{r}}u_{\bm{k_{\mu}}}(\bm{r}),
\end{align}
where the envelope function is $F_{n,\mu}(\bm{r})=\sum_{\bm{k}\in \Gamma_{\mu}}F_{n,\mu}(\bm{k})e^{i (\bm{k}-\bm{k_{\mu}}).\bm{r}}$. 
The implicit-summation PDEs are
\begin{align}\label{eq:fulleig}
(H-W_n)\psi_n(\bm{r})=\zeta(\bm{r}) \psi_{n-1}(\bm{r}),
\end{align}
where $\zeta(\bm{r})=\bm{\epsilon}.\bm{r}$. Substituting $\psi_n(\bm{r})$ from Eqn~\eqref{eq:blochexp}
\begin{align}
\sum_{\mu}\sum_{\bm{k}\in \Gamma_{\mu}}(E(\bm{k})+V(r)-W_n)F_{n,\mu}(\bm{k})\phi_{\bm{k}}(\bm{r})=\sum_{\mu}\sum_{\bm{k}\in \Gamma_{\mu}}\zeta(\bm{r}) F_{n-1,\mu}(\bm{k})\phi_{\bm{k}}(\bm{r}).
\end{align}
Multiply by $\phi^{*}_{\bm{k'}}(\bm{r})$ with $\bm{k}'$ from the valley $\Gamma_{\nu}$, integrate over $\bm{r}$ and use the orthonormality of the Bloch functions, we have
\begin{align}
(E(\bm{k}')-W_{n})F_{n,\nu}(\bm{k}')+\sum_{\mu}\sum_{\bm{k}\in \Gamma_{\mu}}F_{n,\mu}(\bm{k})\int d\bm{r} \, \phi^{*}_{\bm{k'}}(\bm{r})V(r)\phi_{\bm{k}}(\bm{r})= \sum_{\mu}\sum_{\bm{k}\in \Gamma_{\mu}} F_{n-1,\mu}(\bm{k})\int d\bm{r} \, \phi^{*}_{\bm{k'}}(\bm{r})\zeta(\bm{r}) \phi_{\bm{k}}(\bm{r}) 
\end{align}
for $\bm{k'}$ closed to $k_{\nu}$ and $\bm{k}$ closed to $k_{\mu}$, we apply the usual approximation of multivalley EMT and replace the atomic parts $u_{\bm{k'}}(\bm{r})$ and $u_{\bm{k}}(\bm{r})$ in the Bloch functions by $u_{\bm{k_\nu}}(\bm{r})$ and $u_{\bm{k_\mu}}(\bm{r})$, respectively. The above equation becomes
\begin{align}\label{eq:mv1}
(E(\bm{k}')-W_{n})F_{n,\nu}(\bm{k}')&+\sum_{\mu}\int d \bm{r}\,e^{-i(\bm{k'}-\bm{k_{\nu}}).\bm{r}}\left(\sum_{\bm{k}\in \Gamma_{\mu}}F_{n,\mu}(\bm{k})e^{i(\bm{k}-\bm{k_{\mu}}).\bm{r}}\right)V_{\nu \mu}(\bm{r})\nonumber \\
&=\sum_{\mu}\int d\bm{r} \, e^{-i(\bm{k'}-\bm{k_{\nu}}.\bm{r})} \left(\sum_{\bm{k}\in \Gamma_{\mu}} F_{n-1,\mu}(\bm{k}) e^{i(\bm{k}-\bm{k_{\mu}}).\bm{r}}\right)\zeta_{\nu \mu}(\bm{r}),\nonumber \\
(E(\bm{k}')-W_{n})F_{n,\nu}(\bm{k}')&+\sum_{\mu}\int d \bm{r}\,e^{-i(\bm{k'}-\bm{k_{\nu}}).\bm{r}}F_{n,\mu}(\bm{r})V_{\nu \mu}(\bm{r})\nonumber \\
&=\sum_{\mu}\int d\bm{r} \, e^{-i(\bm{k'}-\bm{k_{\nu}}).\bm{r}} F_{n-1,\mu}(\bm{r})\zeta_{\nu \mu}(\bm{r}),
\end{align}
where the valley-orbit coupling potential and dipole operator are
\begin{align}
V_{\nu \mu}(\bm{r})&=\phi^{*}_{\bm{k}_{\nu}}(\bm{r})V(\bm{r})\phi_{\bm{k}_{\mu}}(\bm{r}) \nonumber \\
\zeta_{\nu \mu}&=\phi^{*}_{\bm{k}_{\nu}}(\bm{r})\zeta(\bm{r})\phi_{\bm{k}_{\mu}}(\bm{r}).
\end{align}
Multiply Eqn~\eqref{eq:mv1} by $e^{i(\bm{k'}-\bm{k_\nu}).\bm{r'}}$ and summing over $\bm{k'}\in\Gamma_{\nu}$ we have for each term 
\begin{align}
\sum_{\bm{k}'\in\Gamma_{\nu}}(E(\bm{k}')-W_{n})F_{n,\nu}(\bm{k}')e^{i(\bm{k'}-\bm{k}_{\nu}).\bm{r}'}&=\sum_{\bm{k}'\in\Gamma_{\nu}}(E(-i\bm{\nabla}'+\bm{k}_{\nu})-W_n)F_{n,\nu}(\bm{k}')e^{i(\bm{k'}-\bm{k}_{\nu}).\bm{r}'}\nonumber \\
&=(E(-i\bm{\nabla}'+\bm{k}_{\nu})-W_n)F_{n,\nu}(\bm{r}'), \nonumber \\
\sum_{\bm{k}'\in\Gamma_{\nu}}\sum_{\mu}\int d \bm{r}\,e^{-i(\bm{k'}-\bm{k_{\nu}}).(\bm{r}-\bm{r'})}F_{n,\mu}(\bm{r})V_{\nu \mu}(\bm{r})&=\sum_{\mu}\int d \bm{r}\,\left[\sum_{\bm{k}'\in\Gamma_{\nu}}e^{-i(\bm{k'}-\bm{k_{\nu}}).(\bm{r}-\bm{r'})}\right]F_{n,\mu}(\bm{r})V_{\nu \mu}(\bm{r})\nonumber \\
&\approx\sum_{\mu}\int d \bm{r}\,\delta(\bm{r}-\bm{r'})F_{n,\mu}(\bm{r})V_{\nu \mu}(\bm{r})=\sum_{\mu}F_{n,\mu}(\bm{r}')V_{\nu \mu}(\bm{r}'),\nonumber \\
\sum_{\bm{k}'\in\Gamma_{\nu}}\sum_{\mu}\int d\bm{r} \, e^{-i(\bm{k'}-\bm{k_{\nu}}).(\bm{r}-\bm{r}')} F_{n-1,\mu}(\bm{r})\zeta_{\nu \mu}(\bm{r})&\approx\sum_{\mu}F_{n-1,\mu}(\bm{r'})\zeta_{\nu \mu}(\bm{r'}),
\end{align}
and we arrive at the multivalley coupled equations for the envelope functions
\begin{align}
(E(-i\bm{\nabla}+\bm{k}_{\nu})-W_n)F_{n,\nu}(\bm{r}) +\sum_{\mu}F_{n,\mu}(\bm{r})V_{\nu \mu}(\bm{r})=\sum_{\mu}F_{n-1,\mu}(\bm{r})\zeta_{\nu \mu}(\bm{r}).
\end{align}
Within the parabolic approximation for the conduction band minimum 
\begin{align}
E(-i\bm{\nabla}+\bm{k}_{\nu})\approx -\frac{\hbar^2}{2m_t}\left(\frac{\partial^2}{\partial x_{\nu}^2}+\frac{\partial^2}{\partial y_{\nu}^2}\right)-\frac{\hbar^2}{2 m_l}\frac{\partial^2}{\partial z_{\nu}^2}.
\end{align}
If we further neglect the inter-valley coupling due to the long range part $-e^2/(4\pi\epsilon_0\epsilon_r r)$ of the impurity potential and the inter-valley coupling in the dipole matrix element, and use $\phi^{*}_{\bm{k}_{\nu}}(\bm{r})\phi_{\bm{k}_{\nu}}(\bm{r})\approx 1$ we have

\begin{align}
\left[-\frac{\hbar^2}{2m_t}\left(\frac{\partial^2}{\partial x_{\nu}^2}+\frac{\partial^2}{\partial y_{\nu}^2}\right)-\frac{\hbar^2}{2 m_l}\frac{\partial^2}{\partial z_{\nu}^2}-\frac{e^2}{4\pi\epsilon_0\epsilon_r r}-W_n\right]F_{n,\nu}(\bm{r})+\sum_{\mu}U_{\nu \mu}(\bm{r})F_{n,\mu}(\bm{r})=\zeta(\bm{r})F_{n-1,\mu}(\bm{r}),
\end{align}
where
\begin{align}
U_{\nu \mu}(\bm{r})&=\phi^{*}_{\bm{k}_{\nu}}(\bm{r})U(\bm{r})\phi_{\bm{k}_{\mu}}(\bm{r}). 
\end{align}
If $U(\bm{r})$ is very short range  $U_{\nu \mu}(\bm{r})$ effectively samples the slow-varying envelope function at $r=0$, hence it is a good approximation to replace $U_{\nu \mu}(\bm{r})$ by $U_{\nu \mu}\delta (\bm{r})$ where 
\begin{equation}
U_{\nu \mu}=\int d\bm{r}\phi^{*}_{\bm{k}_{\nu}}(\bm{r})U(\bm{r})\phi_{\bm{k}_{\mu}}(\bm{r}),
\end{equation}
and we arrive at the same equation as Eqn~\eqref{eq:mvpde} in Sec. III with the only difference being the form of $U_{\nu,\mu}$ which now has a contribution from $u_{\bm{k}_{\mu}}(\bm{r})$.

To obtain the multivalley Schrodinger equation discussed in Sec. I we replace in Eqn~\eqref{eq:fulleig} $\ket{\psi_n}$ by $\ket{\psi_j}$, $W_n$ by $\hbar \omega_j$ and set $\zeta (\bm{r})=0$. Repeating the same steps we arrive at 
\begin{align}
\left[-\frac{\hbar^2}{2m_t}\left(\frac{\partial^2}{\partial x_{\nu}^2}+\frac{\partial^2}{\partial y_{\nu}^2}\right)-\frac{\hbar^2}{2 m_l}\frac{\partial^2}{\partial z_{\nu}^2}-\frac{e^2}{4\pi\epsilon_0\epsilon_r r}-\hbar \omega_j\right]F_{j,\nu}(\bm{r})+\sum_{\mu}U_{\nu \mu}\delta (\bm{r}) F_{j,\mu}(\bm{r})=0,
\end{align}  
which is the same as Eqn~\eqref{eq:mvpde} but with a different form for $U_{\nu \mu}$. This is the Shindo-Nara multivalley efffective mass equation of Ref.~\cite{shindo1976}. 

\section{Multivalley N-photon absorption}
The multi photon matrix element is produced by successive applications of the operators ${G}_n$ and $\zeta$, both of which mix $m$ states. Since ${G}_n$ contains ${H}$ which contains $U_{cc}$, it produces inter-valley mixing (while ${H}_0$ and $\zeta$ do not). Consider the excitation from $\ket{\psi_g}\equiv \ket{\psi_0}$ to $\ket{\psi_e}$ 
\begin{eqnarray}
M^{(N)}
&=& \bra{\psi_e}\zeta {G}_{N-1}\zeta \ldots {G}_2 \zeta {G}_1\zeta\ket{\psi_g} \nonumber\\
&=& \bra{\psi_e}\zeta \ket{\psi_{N-1}}. \nonumber 
\end{eqnarray}
Now, often the excited state is part of a degenerate manifold, e.g. in our Si:P situation the states with $m$ and $-m$ are degenerate. Also, the multivalley states with odd parity envelope have zero amplitude at $r=0$, so the CCC can also be neglected and the $A_1$, $E$ and $T_2$ states are degenerate. At the same time, the intermediate state $\zeta\ket{\psi_{N-1}}$ also has multiple components.
Let the set of degenerate components of the excited state be $\psi_e$ which is a subset of the complete set of eigenstates and let the remainder be $\psi_d$, so that we may express the intermediate state as
\[
\zeta\ket{\psi_{N-1}}=\sum_e \alpha_e\ket{\psi_e} +\sum_d \alpha_d\ket{\psi_d} 
\]
where 
$
\alpha_j=\braket{\psi_j|\zeta|\psi_{N-1}}
$. The excited state superposition that has greatest overlap with this state is 
\[
\ket{\Psi_e}= \frac{\sum_e \alpha_e\ket{\psi_e}}
{\sqrt{\sum_e \vert \alpha_e\vert^2}},
\]
and the matrix element is then
\begin{eqnarray}
M^{(N)}&=&\frac{\sum_e \alpha_e\bra{\psi_e}}
{\sqrt{\sum_e \vert \alpha_e\vert^2}}
\sum_e' \alpha_e'\ket{\psi_e'} 
= \sqrt{\sum_e \vert \alpha_e\vert^2}, \nonumber \\
\vert M^{(N)} \vert^2 &=& \sum_e \vert \bra{\psi_e} \zeta\ket {\psi_{N-1}} \vert^2, \label{generalMEsuperposition}
\end{eqnarray}
so the N-photon transition rate is the total transition rate to the degenerate manifold as it should be.

Resolving $\ket{\psi_{N-1}}$ into valley and azimuthal components  (combining Eqns~\eqref{eq:n_expansion_mu} and \eqref{eq:n_expansion_m}
\[
\psi_{N-1}(\bm{r})=\sum_{m,\mu}e^{i\mathbf{k}_\mu .\mathbf{r} } e^{im\phi}f_{N-1,m,\mu}(\bm{r}), 
\]
and similarly for 
\[
\psi_e(\bm{r})=\sum_\mu e^{i\mathbf{k}_\mu .\mathbf{r} } e^{im\phi}f_{e,m,\mu}(\bm{r}), 
\]
and 
\begin{equation}\label{eq:zetacomp}
\zeta=\sum_{\Delta m=-1,0,1} \zeta_{\Delta m,\mu} e^{i\Delta m \phi }. 
\end{equation}
Calculating the matrix element by averaging over a volume $(2\pi/k_0)^3$ and integrating over $\phi$ as usual,
\begin{equation}\label{MEmcomponent}
 \bra{\psi_e} \zeta\ket {\psi_{N-1}} =\sum_{\mu,\Delta m=-1,0,1} \bra{f_{e,m+\Delta m,\mu}}\zeta_ {\Delta m,\mu}\ket {f_{N-1,m,\mu}}.
\end{equation}

\section{Non-linear susceptibility} 
The linear susceptibility (Eqn 3.2.23 in Ref.~\cite{Boyd}) for polarization along axis $x_\nu$ produced by light polarized along $x_{\nu'}$ is
\begin{align}
\chi^{(1)}(\omega)=\frac{n_{3D} e^2}{\epsilon_0\hbar}\sum_{j}\bra{\psi_g}\zeta \ket{j}\bra{j}\zeta \ket{\psi_g} \left[\frac{1}{(\omega_{jg}-\omega)}+\frac{1}{(\omega_{jg}+\omega)} \right],
\end{align}
where $e$ is the electron charge and $n_{3D}$ is the density of atoms, and the frequency $\omega$ is far from any resonances. For an incoming wave with just one polarization and frequency the  second term in the bracket  is ``antiresonant.''
\begin{align}
\chi^{(1)}(\omega)=\frac{n_{3D}e^2}{\epsilon_0\hbar}\sum_{j} \frac{\bra{\psi_g}\zeta\ket{j}\bra{j}\zeta\ket{\psi_g}}{(\omega_{jg}-\omega)}=\frac{n_{3D}e^2a_B^2}{\epsilon_0 E_H} C^{(1)}(\omega),
\end{align}
where
\begin{align}
C^{(1)}(\omega)= \bra{\psi_g} \zeta G_1 \zeta\ket{\psi_g}.
\end{align}
The second order susceptibility is zero because it contains terms like 
$\bra{\psi_g} \zeta G_2 \zeta G_1 \zeta\ket{\psi_g}$
which are forbidden by parity since $\zeta$ is odd. The third order susceptibility is (Eqn 3.2.37 in Ref.~\cite{Boyd})
\begin{equation}\label{eq:fullchi3}
  \chi^{(3)}(3\omega)=\!\begin{aligned}[t]\frac{n_{3D}(ea_B)^4}{\epsilon_0\hbar^3}\sum_{l,k,j}& \bra{\psi_g}\zeta\ket{l}\bra{l}\zeta\ket{k}\bra{k}\zeta \ket{j}\bra{j}\zeta\ket{\psi_g} \\[-1ex]
  & \biggl[ \frac{1}{(\omega_{lg}-3\omega)(\omega_{kg}-2\omega)(\omega_{jg}-\omega)}  
    +\frac{1}{(\omega_{lg}+\omega)(\omega_{kg}-2\omega)(\omega_{jg}-\omega)} \\[-1ex]
  & +\frac{1}{(\omega_{lg}+\omega)(\omega_{kg}+2\omega)(\omega_{jg}-\omega)}  
  +\frac{1}{(\omega_{lg}+\omega)(\omega_{kg}+2\omega)(\omega_{jg}+3\omega)}
  \biggr],
  \end{aligned} \end{equation}
where the first term in the bracket is the resonant term, and the other three are ``antiresonant''.
\begin{equation}
  \chi^{(3)}(3\omega)=\frac{n_{3D}(ea_B)^4}{\epsilon_0 E_H^3} C^{(3)}(3\omega),
 \end{equation}
where 
\begin{equation}
  C^{(3)}(3\omega)= \bra{\psi_g} \zeta G_3 \zeta G_2 \zeta G_1 \zeta\ket{\psi_g} =  \bra{\psi_g} \zeta \ket{\psi_3}, 
 \end{equation}
for the resonant term. The other terms are produced by simple changes in $G_n$.  

The resonant part of the N-th order susceptibility is 
\begin{equation}
  \chi^{(N)}(N\omega)=\frac{n_{3D}(ea_B)^{N+1}}{\epsilon_0 E_H^N} C^{(N)}(N\omega),
 \end{equation}
where
\begin{equation}\label{eq:NsusME}
  C^{(N)}(N\omega)= \bra{\psi_g} \zeta G_n \ldots \zeta {G}_2 \zeta G_1 \zeta\ket{\psi_g} =   \bra{\psi_g} \zeta \ket{N}. 
 \end{equation}
Clearly, extraction of the non-linear susceptibility requires calculation of a  matrix element  that is very similar to that for the multiphoton absorption involving the same PDEs, except that now $\omega\neq \omega_{eg}/N$. The final integral for the matrix element can be made in the same way as for $M^{(N)}$ with Eqn \eqref{MEmcomponent}.

\section{Third harmonic generation}
The susceptibility calculated in the previous section is used for the prediction of non-linear optical processes. Here we give definitions to allow cross-comparison of the strength of one resulting effect, third harmonic generation (3HG) in hydrogenic donors in silicon and hydrogenic atoms in a gas.

External electric fields can polarize a medium, and if it is non-linear this can lead to the appearance of new frequency components, different from the external drive frequency. 
%In other words, time-varying induced nonlinear polarization can acts as a source of new frequency components of the electromagnetic wave. 
Since Si:P has an isotropic potential in an isotropic dielectric, its nonlinear response is determined to lowest order by $\chi^{(3)}$, and the corresponding third order polarization $\vec{P}^{(3)}$ has the form:
\begin{equation}\label{cubic_polarization}
\vec{P}^{(3)}=\epsilon_0 \chi^{(3)}\vec{E}^3,
\end{equation}
where $\epsilon_0$ is the permittivity of free space and the drive  electric field has amplitude  $\vec{E}$ with  single frequency component and propagating along $z$ axis for simplicity 
\begin{equation}\label{electric_field}
E(z,t)=E(z)e^{-i\omega_p t} + \text{c.c.},
\end{equation}
where $E(z)$ is the complex amplitude of the electric field:
\begin{equation}\label{electric_field_amplitude}
E(z)=A_p(z)e^{-i k_p  z}.
\end{equation}
Using Eqn~\eqref{cubic_polarization} one can obtain the polarization induced by the applied electric field 
\begin{equation}\label{cubic_polarization_explicit}
P^{(3)}=\epsilon_0 \chi^{(3)} \left(E(z)e^{-i\omega_p  t}+E^*(z)e^{i\omega_p t}\right)^3=\epsilon_0 \chi^{(3)} E^3(z) e^{-i 3 \omega_p  t} + 3 \epsilon_0 \chi^{(3)} E^2(z) E^*(z) e^{-i \omega_p  t} + \text{c.c.}
\end{equation}
The first term on the right hand side of this equation leads to  third harmonic generation; and one can represent the third order polarization at the triple frequency ($\omega_o=3\omega_p$) as
\begin{equation}\label{cubic_polarization_at_triple_freq}
P_{3\omega_p}^{(3)}(z,t)=\epsilon_0 \chi^{(3)} A_p^3(z) e^{i 3 k_p z} e^{-i 3 \omega_p t}.
\end{equation}

Wave coupling in nonlinear polarized media results in energy transfer between interacting waves. The wave equation for isotropic, nonmagnetic, dielectric media with nonlinear polarization $\vec{P}_{\text{NL}}$ results straightforwardly from the Maxwell equations and has the form:
\begin{equation}\label{wave_equation}
    \nabla^2 \vec{E}=\frac{\varepsilon_r}{c^2}\frac{\partial^2}{\partial t^2}\vec{E}-\frac{1}{\epsilon_0 c^2}\frac{\partial^2}{\partial t^2}\vec{P}_{\text{NL}},
\end{equation}
where $\varepsilon_r$ is the relative dielectric permittivity. In the case of isotropic media $\varepsilon_r$ is a scalar quantity and the (linear) refraction index $n^2=\varepsilon_r$. We suppose that this equation is satisfied for each frequency component separately. 
Let's obtain derivatives of Eqn~\eqref{wave_equation} for the output ($3\omega_p$) frequency component explicitly.
\begin{equation}\label{cubic_polarization_at_triple_freq_d2P/dt2}
\frac{\partial^2}{\partial t^2}P_{3\omega_p}= - \left(3\omega_p\right)^2 \epsilon_0 \chi^{(3)} A_p^3(z) e^{i 3 k_p z} e^{-i 3 \omega_p t},
\end{equation}
\begin{equation}\label{electric_field_d2E/dt2}
\frac{\partial^2}{\partial t^2}E_{3\omega_p}(z,t)= - A_o(z) \omega_o^2 e^{i k_o z} e^{-i \omega_o t},
\end{equation}
\begin{equation}\label{electric_field_d2E/dz2}
\frac{\partial^2}{\partial z^2}E_{3\omega_p}(z,t)= e^{i k_o z} e^{-i \omega_o t} \left( \frac{\partial^2}{\partial z^2}A_o(z) + 2 i k_o \frac{\partial}{\partial z}A_o(z) - A_o(z) k_o^2 \right),
\end{equation}
where we can neglect the first item in the parenthesis as we suppose that the complex amplitude of the field varies slowly along $z$:
\begin{equation*}
k \frac{\partial}{\partial z}A_o(z) \gg \frac{\partial^2}{\partial z^2}A_o(z).
\end{equation*}
Putting Eqns~ \eqref{cubic_polarization_at_triple_freq_d2P/dt2}, \eqref{electric_field_d2E/dt2}, \eqref{electric_field_d2E/dz2} into \eqref{wave_equation} one can get
\begin{equation}\label{wave_equation_for_triple_frequency}
\left(2 i k_o \frac{\partial}{\partial z}A_o(z) - A_o(z) k_o^2 \right) e^{i k_o z} e^{-i \omega_o t} + \frac{\varepsilon_r(3\omega_p)}{c^2} A_o(z) \omega_o^2 e^{i k_o z} e^{-i \omega_o t} = - \frac{1}{c^2} \left(3\omega_p\right)^2 \chi^{(3)} A_p^3(z) e^{i 3 k_p z} e^{-i 3 \omega_p t}
\end{equation}
Inserting $\varepsilon_r/c^2=k^2/\omega^2$, $3\omega_p=\omega_o$ and introducing the wavevector mismatch $\Delta k = 3 k_p - k_o$, Eqn~\eqref{wave_equation_for_triple_frequency} becomes
\begin{equation}\label{amplitude_equation_coupled}
\frac{\partial A_o(z)}{\partial z} = \frac{i}{2}\frac{\omega_o}{c n_o} \chi^{(3)} A_p^3(z) e^{i \Delta k z}.
\end{equation}
This result is the so-called coupled-amplitude equation because it shows how the amplitude of the output wave depends on the three amplitudes of the pumping waves and the phase mismatch between them. It should to be noted that we can deduce similar equations for each of the interacting pump waves, but it is already clear that all amplitudes are functions of each other and there is no simple solution. However, if we  assume that conversion of the pump power into the generated wave is small, i.e. the amplitude of the pumping wave is a constant along the nonlinear medium, then we can integrate Eqn~\eqref{amplitude_equation_coupled}:
\begin{equation}\label{amplitude_integrated}
A_o(z)=\frac{i}{2}\frac{\omega_o}{c n_o} \chi^{(3)} A_p^3 \int_0^L e^{i \Delta k z} \,dz = - \frac{\omega_o}{2 c n_o } \chi^{(3)} A_p^3 \left( \frac{1-e^{i \Delta k L}}{\Delta k} \right),
\end{equation}
where $L$ is the length of nonlinear medium. When $L=L_c = \pi / \Delta k$, the term in the parentheses of Eqn~\eqref{amplitude_integrated} reaches its maximum value of $2/\Delta k$. The length $L_c$ is the so-called the coherence length, and above it  power is converted backwards, from the third harmonic to the pump. Thus it limits the useful length of a media. For normal dispersion $\Delta k > 0$ in isotropic materials,  but fortunately in the THz region a frequency dependence of the refraction is weak enough that the coherent length be as much as some tens centimeters.

Now we are ready to obtain an intensity of the output beam:
\begin{equation}
I_o = 2 n_o \epsilon_0 c A_o A_o^* = 2 n_o \epsilon_0 c \frac{\omega_o^2}{4 c^2 n_o^2 \Delta k^2} \left(\chi^{(3)}\right)^2 A_p^6 \left( 1-e^{i \Delta k L} \right) \left( 1-e^{-i \Delta k L} \right).
\end{equation}
Introducing parameter $x=\Delta k L/2$ the product of the final pair of  brackets is $4 (\sin x)^2$ and finally we get
\begin{equation}\label{output_intensity}
I_o = \frac{\omega_o^2 L^2 I_p^3 }{16 c^4 \epsilon_0^2 n_o n_p^3} \left(\chi^{(3)}\right)^2 \left(\frac{\sin x}{x}\right)^2.
\end{equation}
One can represent this equation over the beam powers $P$ in W and pump frequency $\nu_p$ in Hz:
\begin{equation}\label{output_power}
P_o = \frac{36 \nu_p^2 L^2 P_p^3 }{c^4 \epsilon_0^2 n_o n_p^3 d^4} \left(\chi^{(3)}\right)^2 \left(\frac{\sin x}{x}\right)^2,
\end{equation}
where $d$ is the beam diameter and all parameters have MKS dimensions.This expression for plane waves \cite{Maker1966,Shen} may be modified for 3HG in a gaussian beam \cite{New1967,MH}. Equation~\eqref{output_power} shows that the output power is proportional to the cube of the input power. It is also proportional to the squared length of the nonlinear medium in the  absence of  wave mismatch (i.e. $\Delta k = 0$). It should to be noted that Eqns \eqref{output_intensity}, \eqref{output_power} are valid only for low conversion efficiency ($P_o/P_p \le \approx 1\%$).
%It is so-called phase-matching condition. 
%However due to the intrinsic dispersion, an isotropic media doesn’t satisfy this condition and a constructive interference between of interacted waves will present on so-called coherent length.

The following parameters were used to estimate the 3HG efficiency in Si:P: $n_i=3.41538$ at 4 THz and $n_o=3.41534$ at 12 THz (RT), $\chi^{(3)}=2.9 \cdot 10^{-12}$ (V/m)$^2$ for a donor concentration $N=10^{17}$ cm$^{-3}$. Using $L=1.4$ cm, $d=1$ mm and pump power of 1 W one can get the conversion efficiency of 1\%.

\section{Effect of phonon relaxation and multiphoton ionization}

In this section we discuss procesess that may possibly reduce the 3HG efficiency. Losses due to dephasing by phonon scattering may become important if the time spent in the intermediate states exceeds the phonon lifetime. Since the inverse of the former is given approximately by  the detuning ($\Delta f \Delta t \ge 1/2\pi $) and the inverse phonon-limited width $1/\pi T_2=$ 1 GHz \cite{Steger2009,Greenland2010}, this loss is negligible for much of the spectrum.

Another nonlinear process that may lower the 3HG efficiency is multiphoton ionization \cite{bebb1966} since it reduces the population of the donors in the ground state and creates additional decoherence. When $\omega=\bar{\omega}_{2p}/3$ for example, a four photon absorption takes the electron to the continuum. Here we estimate this ionization in Si:P using the implicit summation method.

The N-photon ionization rate can be estimated if the wavefunction of the ejected electron in the continuum is known \cite{bebb1966}. We use the following approximation for the continuum wavefunction in Si:P: In the stretched frame given by the transformation of Eqn~\eqref{eq:strframe} we replace the asymmetric potential $1/r'\sqrt{1-(1-\gamma)(\cos \theta')^2}$ by its angular average $\kappa/r'$ where 
\begin{equation}
\kappa=\frac{1}{2}\int_0^{\pi}\frac{\sin \theta' d\,\theta'}{\sqrt{1-(1-\gamma)(\cos \theta')^2}}=\frac{\arctan\left(\sqrt{1/\gamma-1}\right)}{\sqrt{1-\gamma}},
\end{equation}
so that the potential is spherical in the stretched frame. Assuming that the energy of the electron in the continuum is within a few tens meV from the conduction band minima, we can again employ effective mass theory. The envelope function of an ejected electron with  wavenumber $\bm{k}'$ in the stretched frame in valley $\mu$ is given by $\Psi_{\mu}(\bm{r'})=e^{i \bm{k'}_{\mu}.\bm{r'}}\sum_{l=0}^{\infty} F_{\bm{k'},l,\mu}(\bm{r'})$ where
\begin{align}
F_{\bm{k'},l,\mu}(\bm{r'})= 4\pi e^{i \alpha_l}\sum_{m=-l}^{l}R_{k',l}(r')Y_{l}^{m}(\theta',\phi')Y_l^{m^*}(\Omega_{k'}),
\end{align} 
where $k'=\sqrt{2m_t E}/\hbar=\sqrt{2E/E_H}/a_B$ with $E$ the energy of the ejected electron, $\Omega_{k'}$ the solid angle of the ejected electron in the stretched frame, $\alpha_l=l \pi/2+\arg[\Gamma(l+1-i\kappa/k' a_B)]$, and $R_{k',l}(r')$ is the radial wavefunction 
\begin{equation}
R_{k',l}(r')=e^{\pi\kappa/2k'\!a_B}\left \vert\frac{\Gamma(l+1-\kappa/k' a_B)}{(2l+1)!}\right \vert(2 k' r')^l e^{-i k' r'}  F(i \kappa/k' a_B + l + 1,2l+2,2 i k' r'),
\end{equation}
where $F$ is the confluent hypergeometric function of the first kind. These functions are taken from Ref.~\cite{bebb1966} and scaled to account for the factor $\kappa$.

The total transiton rate for multiphoton ionization in hydrogen is derived in Ref.~\cite{bebb1966}. Generalizing it for Si:P, taking into account the valley degeneracy, we obtain the total rate from the $1\text{sA}_1$ ground state to the continnuum manifold with energy $E$ 
\begin{equation}
w^{(N)}=\frac{1}{(2\pi)^2} \left(\frac{2\pi\alpha_{\text{fs}}}{\sqrt{\epsilon_r}}\right)^N \left(\frac{I_m}{I_a}\right)^N \frac{E_H}{\hbar} |K^{(N)}|^2 (k' a_B), 
\end{equation}
where  $K^{(N)}$ is a dimensionless matrix element given by
\begin{equation}
|K_{fg}^{(N)}|^2=(1/a_B^3)\sum_{\mu=1}^{6}\int d \Omega_{k'} \left\vert\sum_{l=0}^{\infty}\bra{F_{\bm{k}',l,\mu}}\zeta\ket{F_{N-1,\mu}}\right\vert^2=\frac{(4\pi)^2}{a_B^3}\sum_{\mu=1}^{6}\sum_{l=0}^{\infty}\sum_{m=-l}^{l} \left\vert \sum_{\Delta m=0,\pm 1}\bra{f_{k',l,m}}\zeta_{\Delta m}\ket{f_{N-1,m-\Delta m,\mu}}\right \vert ^2,
\end{equation}
where $F_{N-1,\mu}$ are the envelope functions of the $(N-1)$th intermediate state defined in Eqn \eqref{eq:n_expansion_mu},$f_{N-1,m,\mu}$ its component in Eqn~\eqref{eq:n_expansion_m}, $\zeta_{\Delta m}$ the component of the polarization operator defined in Eqn~\eqref{eq:zetacomp}, and $\ket{f_{k',l,m}}$ are the 2D envelope functions
\begin{equation}
f_{k',l,m}(\theta',\phi')=R_{k',l}(r')\sqrt{\frac{(2l+1)}{4\pi}\frac{(l-m)!}{(l+m)!}}P_l^m(\cos \theta'),
\end{equation}
where $P_l^m$ are the associated Legendre polynomials.

For the $3\omega$ transition to the average frequency of the $2p_0$ and $2p_{\pm}$ level in Si:P, a $4\omega$ absorption takes the electron to the continuum with the energy $E\approx 0.237 E_H=9.46$ meV. We find that, for a sample with  $n_{\text{3D}}=5\times 10^{15}$ $\text{cm}^{-3}$, the ionization rate is $w=3.17 $ $\text{s}^{-1}$ for $I_m=10$  $\text{kW/cm}^2$. This simply means that the pulses must be kept significantly shorter than a second to avoid significant ionization.

\bibliography{chi3bib}